\documentclass[reprint,superscriptaddress,prl]{revtex4-1}
\usepackage{amsmath}
\usepackage{amssymb}
\usepackage{bm}
\usepackage{braket}
\usepackage{color}
\usepackage[varg]{txfonts}
\usepackage{varwidth}
\usepackage{dcolumn}
\usepackage[breaklinks,colorlinks=true,linkcolor=blue,urlcolor=cyan,citecolor=blue]{hyperref}
\usepackage{here}
\usepackage{graphicx}
\usepackage{lettrine}

\begin{document}

\title{Signatures of a magnetic superstructure phase induced by ultrahigh magnetic fields in a breathing pyrochlore antiferromagnet}

\author{M.~Gen}
\email{masaki.gen@riken.jp}
\affiliation{Institute for Solid State Physics, University of Tokyo, Kashiwa 277-8581, Japan}
\affiliation{RIKEN Center for Emergent Matter Science (CEMS), Wako 351-0198, Japan}

\author{A.~Ikeda}
\affiliation{Institute for Solid State Physics, University of Tokyo, Kashiwa 277-8581, Japan}
\affiliation{Department of Engineering Science, University of Electro-Communications, Chofu, Tokyo 182-8585, Japan}

\author{K.~Aoyama}
\affiliation{Department of Earth and Space Science, Graduate School of Science, Osaka University, Osaka 560-0043, Japan}

\author{H.~O.~Jeschke}
\affiliation{Research Institute for Interdisciplinary Science, Okayama University, Okayama 700-8530, Japan}

\author{Y.~Ishii}
\affiliation{Institute for Solid State Physics, University of Tokyo, Kashiwa 277-8581, Japan}

\author{H.~Ishikawa}
\affiliation{Institute for Solid State Physics, University of Tokyo, Kashiwa 277-8581, Japan}

\author{T.~Yajima}
\affiliation{Institute for Solid State Physics, University of Tokyo, Kashiwa 277-8581, Japan}

\author{Y.~Okamoto}
\affiliation{Institute for Solid State Physics, University of Tokyo, Kashiwa 277-8581, Japan}

\author{X.-G. Zhou}
\affiliation{Institute for Solid State Physics, University of Tokyo, Kashiwa 277-8581, Japan}

\author{D.~Nakamura}
\affiliation{Institute for Solid State Physics, University of Tokyo, Kashiwa 277-8581, Japan}
\affiliation{RIKEN Center for Emergent Matter Science (CEMS), Wako 351-0198, Japan}

\author{S.~Takeyama}
\affiliation{Institute for Solid State Physics, University of Tokyo, Kashiwa 277-8581, Japan}

\author{K.~Kindo}
\affiliation{Institute for Solid State Physics, University of Tokyo, Kashiwa 277-8581, Japan}

\author{Y.~H.~Matsuda}
\affiliation{Institute for Solid State Physics,  University of Tokyo, Kashiwa 277-8581, Japan}

\author{Y.~Kohama}
\email{ykohama@issp.u-tokyo.ac.jp}
\affiliation{Institute for Solid State Physics, University of Tokyo, Kashiwa 277-8581, Japan}

\begin{abstract}

The mutual coupling of spin and lattice degrees of freedom is ubiquitous in magnetic materials and potentially creates exotic magnetic states in response to the external magnetic field.
Particularly, geometrically frustrated magnets serve as a fertile playground for realizing magnetic superstructure phases.
Here, we observe an unconventional two-step magnetostructural transition prior to a half-magnetization plateau in a breathing pyrochlore chromium spinel by means of state-of-the-art magnetization and magnetostriction measurements in ultrahigh magnetic fields available up to 600~T.
Considering a microscopic magnetoelastic theory, the intermediate-field phase can be assigned to a magnetic superstructure with a three-dimensional periodic array of 3-up-1-down and canted 2-up-2-down spin molecules.
We attribute the emergence of the magnetic superstructure to a unique combination of the strong spin-lattice coupling and large breathing anisotropy.

\end{abstract}

\date{\today}
\maketitle
{\bf Key words}\\
Frustrated magnetism, Breathing anisotropy, Spin-lattice coupling, Ultrahigh magnetic fields\\

{\bf Significance}\\
The search for exotic magnetic states, such as a quantum spin liquid and a magnetic superstructure, in geometrically frustrated magnets has been a central research topic in the recent 30 years. 
Theoretically, the mutual coupling of spin and lattice degrees of freedom has been proposed to induce rich magnetic phases in various spin models.
Here, we observe an unconventional multi-step magnetostructural transition in a breathing pyrochlore antiferromagnet in ultrahigh magnetic fields, signaling the emergence of a magnetic superstructure phase with a periodic array of collinear 3-up-1-down and canted antiferromagnetic states.
This finding could be attributed to the interplay between the spin-lattice coupling and breathing anisotropy.\\

Superstructures in crystalline solids, where the unit cell of atomic arrangements or electronic states is an integer multiple of the original primitive cell of the lattice, have received considerable attention because of their complexity as well as their association with exotic physical properties.
Prominent examples are surface superstructures such as the Si(111)--$7 \times 7$ state \cite{1959_Sch}, charge/orbital ordering in perovskite manganites \cite{2006_Tok}, and charge density wave states in van der Waals transition-metal dichalcogenides \cite{1975_Wil} and topological kagome superconductors \cite{2021_Li}.
For frustrated spin systems, a variety of quantum-entangled magnetic superstructures can appear in the external magnetic field.
Of particular interest are a series of magnon crystals in a spin-1/2 kagome Heisenberg antiferromagnet \cite{2002_Sch} and successive transformations of singlet-triplet superstructures in a spin-1/2 orthogonal-dimer Heisenberg antiferromagnet \cite{2002_Kod}.
These phenomena are accompanied by a cascade of fractional magnetization plateaus \cite{2013_Nis, 2019_Oku, 1999_Kag, 2012_Jai, 2013_Mat}.

\begin{figure}[b]
\centering
\includegraphics[width=\linewidth]{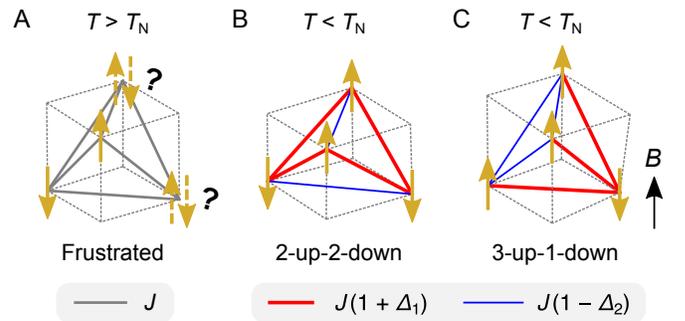}
\caption{Spin configurations in a local tetrahedron of the classical Heisenberg antiferromagnet on a pyrochlore lattice. The magnetic frustration arising from the geometrical constraint (A) is relieved by the spin Jahn-Teller effect below the ordering temperature $T_{\rm N}$ in zero field, resulting in a 2-up-2-down state (B). On application of a magnetic field $B$, the system undergoes a phase transition to a 3-up-1-down state (C). In both collinear magnetic states, the AFM exchange interaction in contracted (elongated) bonds colored by red (blue) becomes stronger (weaker) by $J\Delta_{1}$ ($J\Delta_{2}$) than the original strength $J$ through the lattice deformation \cite{2004_Pen}.}
\label{Fig1}
\end{figure}

When a spin system interacts with the lattice degrees of freedom, the structural instability can facilitate the formation of a magnetic long-range order, potentially yielding a spin-lattice-coupled superstructure.
The concept of a spin-lattice coupling was first theoretically proposed as the spin-Peierls transition in a spin-1/2 Heisenberg antiferromagnetic (AFM) chain \cite{1966_Che, 1971_Pin}, which was then demonstrated in a number of quasi-one-dimensional compounds \cite{1975_Bra, 1993_Has}.
In this mechanism, the spin dimerization is induced by individual atomic displacements with a doubling of the unit cell, where the system gains exchange energy at the cost of elastic energy.
A similar phase transition was subsequently found in three-dimensional semiclassical spin systems represented by chromium-based spinels {\it A}Cr$_{2}$O$_{4}$ ({\it A} = Zn, Cd, and Hg) \cite{2000_Lee, 2002_Rov, 2005_Chu, 2007_Mat}.
At low temperatures, the strong geometrical frustration inherent in the corner-sharing tetrahedral network, i.e., the pyrochlore lattice, of spin-3/2 Cr$^{3+}$ ions is relieved by the local tetrahedral distortion (called the spin Jahn-Teller effect), resulting in a 2-up-2-down magnetic long-range order \cite{2000_Lee, 2007_Mat} (Fig.~\ref{Fig1}A, B).
Interestingly, the application of a magnetic field induces another spin-lattice-coupled long-range order, a 3-up-1-down state \cite{2007_Mat, 2010_Mat} (Fig.~\ref{Fig1}C).
This phase transition is accompanied by a steep magnetization jump \cite{2011_Kim, 2013_Miy, 2011_Miy} and giant magnetostriction \cite{2019_Ros} followed by a robust half-magnetization plateau, which can be explained in the framework of microscopic magnetoelastic theories incorporating local phonon modes \cite{2004_Pen, 2006_Ber}.
These findings have triggered further studies on the spin-lattice-coupling physics for various spin models \cite{2008_Wan, 2013_Alb, 2019_Aoy, 2020_Miy, 2020_Gen, 2021_Aoy, 2022_Gen}.

\begin{figure}[t]
\centering
\includegraphics[width=\linewidth]{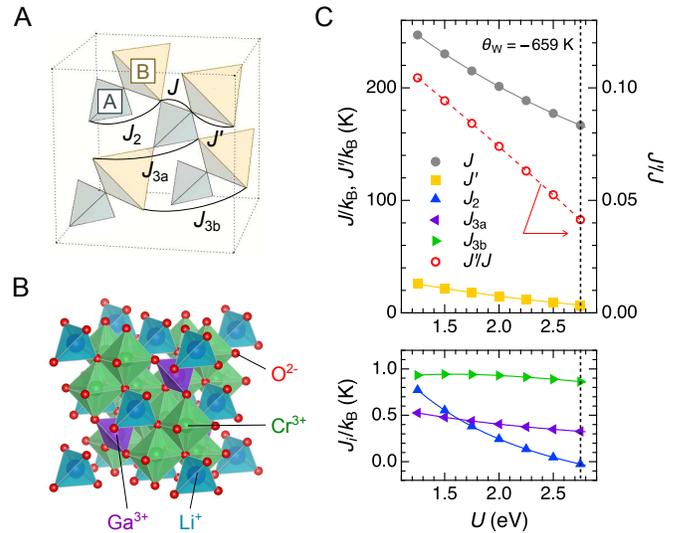}
\caption{Basic properties of a breathing pyrochlore antiferromagnet LiGaCr$_{4}$O$_{8}$. (A) Breathing pyrochlore lattice with two kinds of the nearest-neighbor AFM exchange couplings of $J$ ($J'$) in tetrahedra {\it A} ({\it B}) with the bond length $r$ ($r'$) and further-neighbor exchange couplings up to third nearest-neighbor. In LiGaCr$_{4}$O$_{8}$, $r = 2.907$~$\AA$ and $r' = 2.919$~$\AA$ at 20~K. (B) Crystal structure of LiGaCr$_{4}$O$_{8}$. The illustrations are drawn with VESTA software \cite{2011_Mom}. (C) Exchange couplings of LiGaCr$_{4}$O$_{8}$ in the paramagnetic state at 20~K determined by DFT energy mapping as a function of on-site interaction strength $U$. The vertical line indicates the $U$ value where the exchange couplings match the experimental Weiss temperature $\Theta_{\rm W} = - 659$~K \cite{2013_Oka}. The resulting exchange parameters are $J/k_{\rm B} = 166.6(2)$~K, $J'/k_{\rm B} = 6.9(2)$~K, $J_{2}/k_{\rm B} = 0.0(1)$~K, $J_{3a}/k_{\rm B} = 0.3(1)$~K, and $J_{3b}/k_{\rm B} = 0.9(1)$~K.}
\label{Fig2}
\end{figure}

Recently, the introduction of a {\it breathing} anisotropy, i.e., a spatial modulation of magnetic interactions, has been recognized as a new approach to control the ground state of various spin models \cite{2017_Tsu, 2017_Ora, 2013_Oka}.
In this context, the emergence of spin-lattice-coupled superstructures in addition to the conventional 2-up-2-down and 3-up-1-down states (not superstructures) are theoretically proposed in the breathing pyrochlore lattice where the neighboring tetrahedra differ in size in an alternating pattern (Fig.~\ref{Fig2}A) \cite{2021_Aoy}.
Here, we report the possible realization of this theoretical prediction in a model compound of the breathing pyrochlore antiferromagnet, LiGaCr$_{4}$O$_{8}$ \cite{2013_Oka, 2016_Lee, 2016_Sah, 2020_Kan, 2021_Pok}.
By means of state-of-the-art magnetization and magnetostriction measurement techniques available in ultrahigh magnetic fields of up to 600~T, we demonstrate that LiGaCr$_{4}$O$_{8}$ exhibits a two-step magnetostructural transition in an intermediate field range between 150~T and 200 T, followed by a half-magnetization plateau up to $\sim$420~T.
The effective spin Hamiltonian for LiGaCr$_{4}$O$_{8}$ is established based on density-functional-theory (DFT) calculations and classical Monte Carlo (MC) simulations.
We show that the combination of the strong spin-lattice coupling and large breathing anisotropy can be responsible for stabilizing the intermediate-field phase, which we assign to a tetrahedron-based superstructure with a three-dimensional periodic array of 3-up-1-down and canted 2-up-2-down spin molecules.

\section*{Results}

\vspace{-0.2cm}
\subsection*{Breathing pyrochlore antiferromagnet LiGaCr$_{4}$O$_{8}$}
\vspace{-0.2cm}

We first introduce the basic properties of the present target LiGaCr$_{4}$O$_{8}$ \cite{2013_Oka}.
The crystallographic ordering of nonmagnetic cations Li$^{+}$ and Ga$^{3+}$ identical to the zinc-blende structure (Fig.~\ref{Fig2}B) leads to the breathing pyrochlore Cr network, as shown in Fig.~\ref{Fig2}A.
The resultant two inequivalent nearest-neighbor Cr--Cr bonds, whose lengths are $r = 2.970$~$\AA$ and $r' = 2.867$~$\AA$ at room temperature \cite{2013_Oka}, are characterized by two distinct AFM exchange couplings $J$ and $J'$:
\begin{equation}
\begin{split}
\label{eq:H_0}
{\mathcal{H}_{0}}=J\sum_{\langle i, j\rangle_{A}}{\mathbf S}_{i} \cdot {\mathbf S}_{j}+J'\sum_{\langle i, j\rangle_{B}}{\mathbf S}_{i} \cdot {\mathbf S}_{j},
\end{split}
\end{equation}
where ${\mathbf S}_{i}$ and ${\mathbf S}_{j}$ denote the classical Heisenberg spins, and $\langle i,j \rangle_{A}$ and $\langle i,j \rangle_{B}$ stand for nearest-neighbor sites within tetrahedra {\it A} and {\it B}, respectively (Fig.~\ref{Fig2}A).
Note that $J$ is equal to $J'$ for the conventional Cr spinel {\it A}Cr$_{2}$O$_{4}$.
For LiGaCr$_{4}$O$_{8}$, $J'$ was initially believed to be stronger than $J$ \cite{2013_Oka} judging from the empirical relationship between the strength of the AFM exchange coupling and the lattice constant (equivalently, the Cr--Cr bond length) in {\it A}Cr$_{2}$O$_{4}$ \cite{2008_Yar}, though the actual situation does not seem so simple.
From the structural point of view, the inversion symmetry breaking at the local Cr site would give rise to the anisotropic deformation of Cr 3$d^{3}$ orbitals.
Indeed, previous DFT energy mapping revealed the dominance of $J$ at room temperature: $J/k_{\rm B} = 100$~K and $J'/k_{\rm B} = 66.2$~K \cite{2019_Gho}, $k_{\rm B}$ being the Boltzmann constant.
Furthermore, the breathing anisotropy $J'/J$ can be strongly temperature dependent as exemplified in a related compound LiInCr$_{4}$O$_{8}$ \cite{2019_Gho}.

\begin{figure*}[t]
\centering
\includegraphics[width=\linewidth]{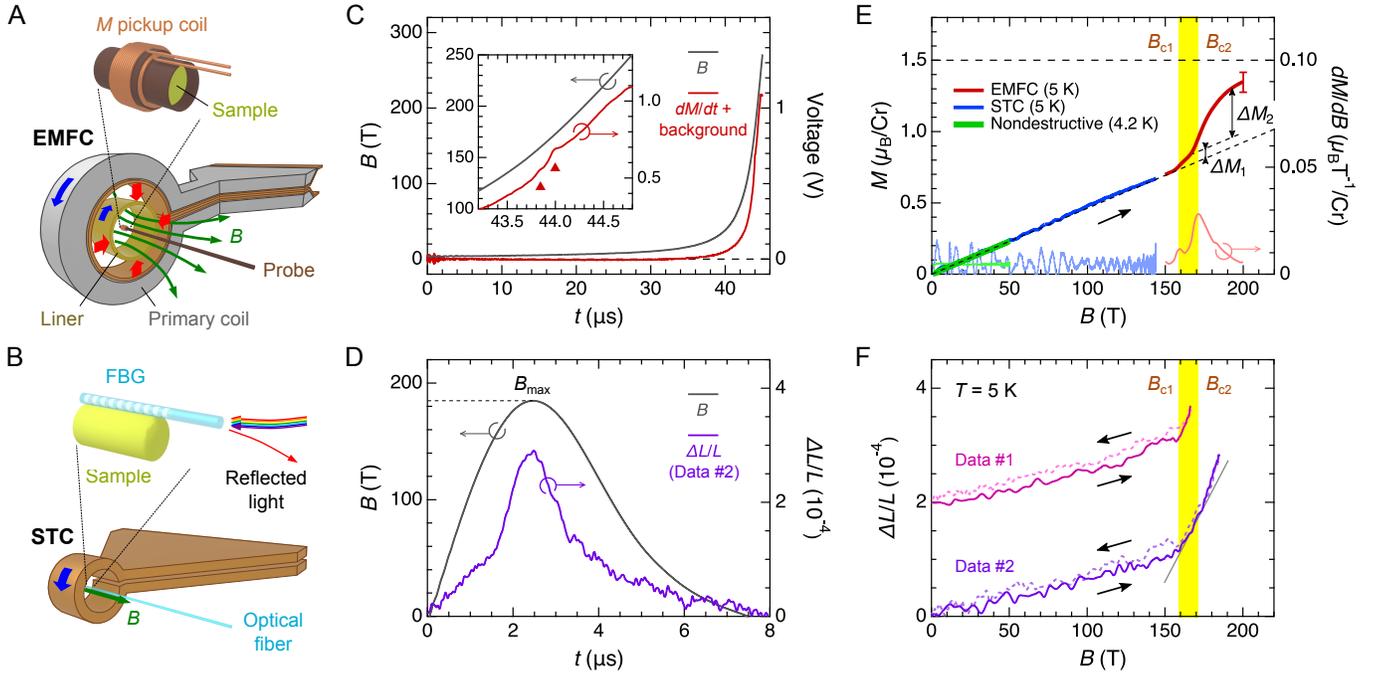}
\caption{Observation of a two-step magnetostructural transition in LiGaCr$_{4}$O$_{8}$ at around 200~T. (A) Schematic of the magnetization measurement using the induction method in the electromagnetic flux compression (EMFC) system. An instantaneous high-current discharge into the outer primary coil induces explosive implosion of the inner brass cylinder (noted as ``Liner'') and as a result compresses the magnetic flux \cite{2018_Nak}. (B) Schematic of the magnetostriction measurement using the fiber-Bragg-grating (FBG) method in the single-turn coil (STC) system. The probe and sample remain intact during the field pulse due to outward explosion of the STC \cite{2003_Miu}. (C) Temporal evolutions of the magnetic field (gray) and the induction voltage detected by the self-compensated pickup coil (red) at 5 K in the EMFC system. The induced voltage is a summation of the intrinsic $dM/dt$ component and the background originating from $dB/dt$. The inset displays an enlarged view around phase transitions denoted by upward triangles in the $dM/dt$ data. (D) Temporal evolutions of the magnetic field (gray) and the sample-length change $\Delta L/L$ (purple) obtained in the STC system (corresponding to Data \#2 in (F)). (E) $M$--$B$ curves (left axis) and their field-derivatives (right axis) at $\sim$5 K obtained in the non-destructive pulsed magnet (green), STC (blue), and EMFC system (red). The parallel dashed lines are guides to the eye to visualize magnitudes of magnetization jumps $\Delta M_{1}$ and $\Delta M_{2}$ at $B_{\rm c1}$ and $B_{\rm c2}$, respectively. (F) $\Delta L/L$--$B$ curves at 5 K obtained in the STC system. Data \#1 and \#2 were taken in the same sample setting with different generated maximum field $B_{\rm max}$ (Data \#1 is vertically shifted for clarity). The gray solid line is a guide for the eye to visualize the slope change in $\Delta L/L$ at $B_{\rm c1}$ and $B_{\rm c2}$.}
\label{Fig3}
\end{figure*}

In order to evaluate $J'/J$ for LiGaCr$_{4}$O$_{8}$ at low temperatures, we measured powder X-ray diffraction patterns at 20~K, and performed the Rietveld analysis (Supplementary Note 1) and DFT energy mapping (Methods).
Since the coexistence of tetragonal and cubic phases are reported below the ordering temperature $T_{\rm N} = 14$~K \cite{2016_Sah} (Supplementary Note 2), we can expect reliable information on the exchange parameters from the structural refinement at a temperature slightly higher than $T_{\rm N}$.
Notably, tetrahedra {\it A} and {\it B} are found to be reversed in size ($r = 2.907$~$\AA$ and $r' = 2.919$~$\AA$) while the cubic $F{\overline 4}3m$ structure is preserved at 20 K.
Figure~\ref{Fig2}C shows the DFT energy mapping up to the third nearest-neighbor exchange couplings (Fig.~\ref{Fig2}A).
Here, the experimental Weiss temperature $\Theta_{\rm W} = - 659$~K \cite{2013_Oka} is used to fix the on-site interaction strength $U$.
We find a significant reduction of $J'/J$ on cooling: $J'/J = 0.66$ at room temperature \cite{2019_Gho} whereas $J'/J$ = 0.04 at 20~K ($J/k_{\rm B} = 166.6$~K and $J'/k_{\rm B} = 6.9$~K), indicating that LiGaCr$_{4}$O$_{8}$ is located close to the limit of decoupled AFM tetrahedra at low temperatures.
The estimation of $J'/J$ is nearly independent of $U$ ($J'/J = 0.04 \sim 0.10$ for $U = 1.25 \sim 2.75$~eV), and the second and third nearest-neighbor exchange couplings are negligibly weak.

\vspace{-0.2cm}
\subsection*{Physical property measurements in ultrahigh magnetic fields}
\vspace{-0.2cm}

As suggested from the large negative Weiss temperature $\Theta_{\rm W} = - 659$~K \cite{2013_Oka}, ultrahigh magnetic fields of several hundreds of tesla are required to exceed the strong AFM exchange interactions in LiGaCr$_{4}$O$_{8}$.
Accordingly, we employed two kinds of destructive-type pulsed megagauss generators: an electromagnetic flux compression (EMFC) system \cite{2018_Nak} (Fig.~\ref{Fig3}A) and a single-turn coil (STC) system \cite{2003_Miu} (Fig.~\ref{Fig3}B).
The former allows physical property measurements up to $\sim$600~T only in the field-increasing process, whereas the latter offers the data up to $\sim$200~T both in the field-increasing and decreasing processes.
Typical magnetic-field waveforms generated by the EMFC and STC systems are shown by gray lines in Fig.~\ref{Fig3}C and D, respectively.

Physical property measurements on magnetic insulators in the EMFC system was so far limited to the Faraday rotation and optical absorption measurements \cite{2011_Miy}.
These magneto-optical detection schemes, however, could not be applied to the present polycrystalline LiGaCr$_{4}$O$_{8}$ samples.
In this work, we have extended two existing techniques implemented in the STC system to the EMFC system to observe successive phase transitions of LiGaCr$_{4}$O$_{8}$ up to $\sim$600~T: (i) a magnetization measurement using the induction method \cite{2012_Tak, 2019_Gen, 2020_Gen} and (ii) a magnetostriction measurement using the optical fiber-Bragg-grating (FBG) method \cite{2017_Ike, 2023_Ike} (Methods and Supplementary Notes 3 and 6).

\vspace{-0.2cm}
\subsection*{Two-step magnetostructural transition at around 200~T}
\vspace{-0.2cm}

Figure~\ref{Fig3}E summarizes the magnetization data of LiGaCr$_{4}$O$_{8}$ measured at $\sim$5~K.
In the STC system, a linear increase in the magnetization $M$ with respect to the external magnetic field $B$ is observed up to a generated maximum field $B_{\rm max}$ of 145~T, where $M$ reaches 0.70~$\mu_{\rm B}$/Cr, suggesting that spins are smoothly canting from the 2-up-2-down AFM ground state \cite{2016_Sah}.
Here, the absolute value of $M$ is calibrated using the magnetization data obtained in a SQUID magnetometer MPMS up to 7~T (Supplementary Note 2) and in a non-destructive pulsed magnet up to 51~T (Fig.~\ref{Fig3}E).
Upon the application of a higher field using the EMFC system, $dM/dt$ anomalies with a small hump and a subsequent large hump are observed at $B_{\rm c1}$ = 159~T and $B_{\rm c2}$ = 171~T, respectively (inset of Fig.~\ref{Fig3}C), indicating a two-step metamagnetic transition.
We ensure the reproducibility in three independent experiments with different setups (Supplementary Note 4).
By subtracting the background component from the observed $dM/dt$ profile (Supplementary Note 5), $M$ as a function of $B$ and its field derivative $dM/dB$ between 150~T and 200~T are obtained as shown by red and pink lines, respectively, in Fig.~\ref{Fig3}E.
The second magnetization jump, $\Delta M_{2}$, is approximately six times larger than the first one, $\Delta M_{1}$, if the $B$-linear component is subtracted (Fig.~\ref{Fig3}E).
The absolute value of $M$ at 200~T is $1.35 \pm 0.07$~$\mu_{\rm B}$/Cr, which is close to half the expected saturation magnetization of $\sim$3~$\mu_{\rm B}$/Cr given that the $g$-value is estimated to be $g = 1.98 \sim 2.08$ \cite{2013_Oka, 2016_Lee}.
The present observation is most likely ascribed to the appearance of a half-magnetization plateau as observed in {\it A}Cr$_{2}$O$_{4}$ \cite{2011_Kim, 2013_Miy, 2011_Miy}, given that the magnetization jump is blunted by thermal fluctuations or crystallographic disorders.

\begin{figure}[t]
\centering
\includegraphics[width=0.8\linewidth]{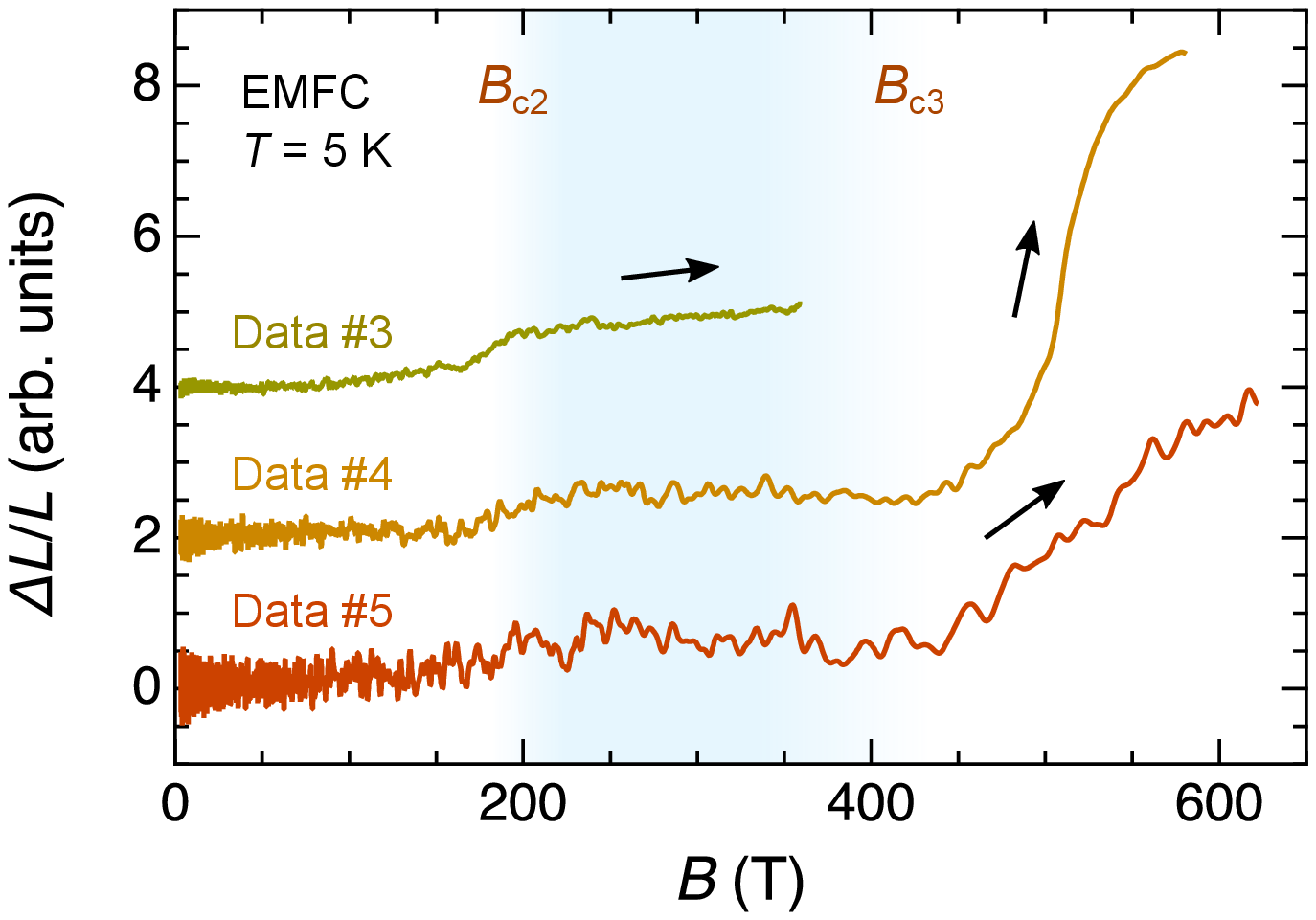}
\caption{Magnetic-field width of a half-magnetization plateau in LiGaCr$_{4}$O$_{8}$ determined by the magnetostriction measurements. $\Delta L/L$--$B$ curves at 5~K obtained in the electromagnetic flux compression (EMFC) system. Data \#3, \#4, and \#5 were taken in independent sample settings with the generated maximum field of $B_{\rm max} = 360$, 580, and 620~T, respectively (Data \#3 and \#4 are vertically shifted for clarity). The half-magnetization plateau associated with the 3-up-1-down magnetic state is expected to appear in a field region shaded by cyan.}
\label{Fig4}
\end{figure}

The existence of an intermediate-field phase between $B_{\rm c1}$ and $B_{\rm c2}$ is supported by the magnetostriction measurements.
Figure~\ref{Fig3}F shows the magnetostriction data of LiGaCr$_{4}$O$_{8}$ measured at 5~K using the STC system with $B_{\rm max} = 167$~T (Data \#1) and 185 T (Data \#2).
The temporal evolution of the relative sample-length change $\Delta L/L$ for Data \#2 is shown in Fig.~\ref{Fig3}D.
Data \#1 and \#2 were taken in the same setup and are in excellent agreement with each other.
As seen in Data \#1, a rapid lattice expansion starts at around 160~T, which corresponds to the phase transition at $B_{\rm c1}$.
Furthermore, as seen in Data \#2, the increase in $\Delta L/L$ is accelerated above $B_{\rm c2}$.
These observations indicate that both transitions at $B_{\rm c1}$ and $B_{\rm c2}$ are likely first order accompanied by a structural phase transition.

\vspace{-0.2cm}
\subsection*{Field width of half-magnetization plateau}
\vspace{-0.2cm}

Theoretically, the magnetic-field width of a half-magnetization plateau correlates with the strength of the spin- lattice coupling in the (breathing) pyrochlore-lattice Heisenberg antiferromagnet \cite{2004_Pen, 2021_Aoy}.
Therefore, it is important to determine the termination field of the half-magnetization plateau $B_{\rm c3}$ for further discussions, although our magnetization measurement could not detect additional phase transitions above 200~T due to poor sensitivity of our probe in the high-field region (Supplementary Note 4).

As an alternative way to determine $B_{\rm c3}$, we measured the magnetostriction of LiGaCr$_{4}$O$_{8}$ at 5~K in the EMFC system.
Considering the relation $\Delta L/L \propto M^{2}$ derived from the magnetoelastic theory \cite{2019_Ros}, there should be no lattice-constant change in the half-magnetization plateau region, followed by a significant volume expansion on the high-field side.
Figure~\ref{Fig4} shows three $\Delta L/L$--$B$ curves (Data \#3 $\sim$ \#5) obtained in independent sample settings.
The raw data are shown in Supplementary Note 7, and the analytical procedure is described in Supplementary Note 8.
For Data \#3, a plateau-like behavior is observed from 200~T up to $B_{\rm max} = 360$~T.
For Data \#4 and \#5 with $B_{\rm max} \approx 600$~T, on the other hand, an upturn behavior is clearly observed at around 420~T, signaling the occurrence of a phase transition from the 3-up-1-down to a higher-field spin-canted phase.
We hence determine $B_{\rm c3} \approx 420$~T.
Furthermore, another $\Delta L/L$ kink is observed at around 550~T, potentially reflecting a phase transition to a paramagnetic phase.

\begin{figure*}[t]
\centering
\includegraphics[width=0.9\linewidth]{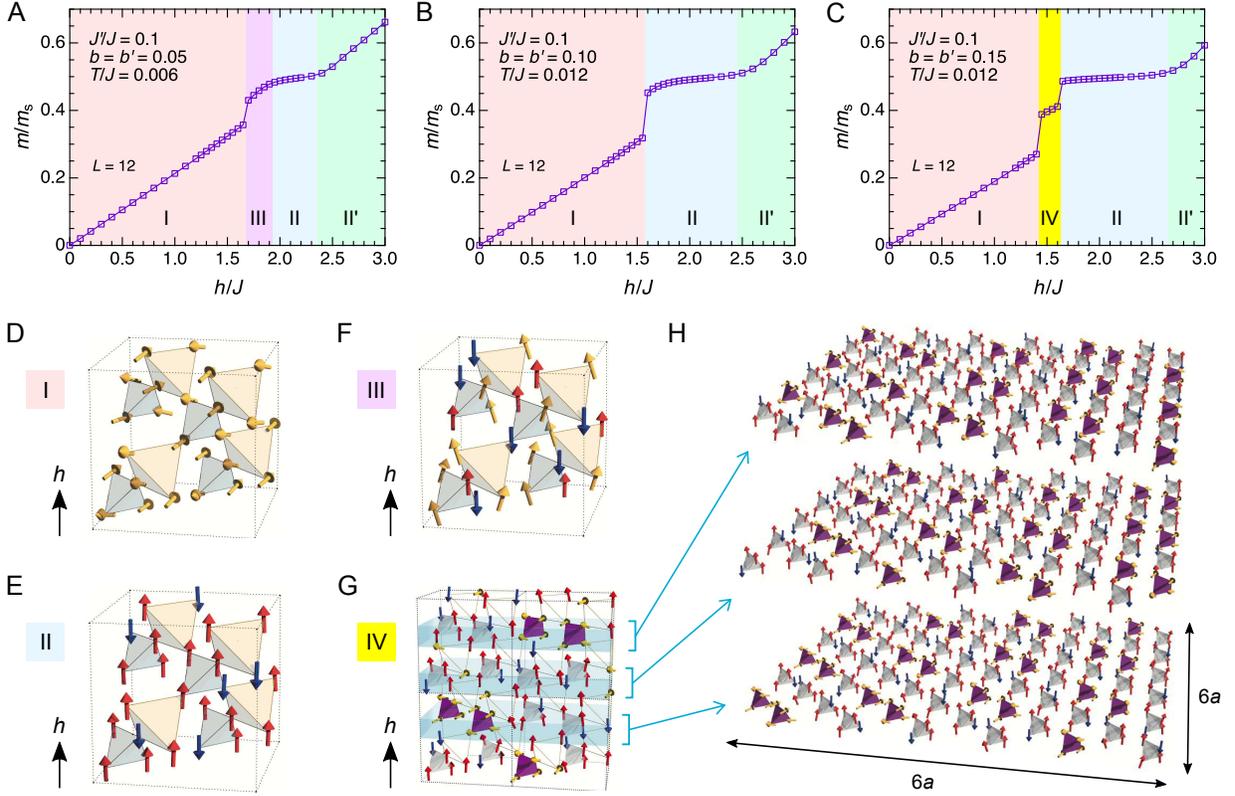}
\caption{Successive field-induced phase transitions based on the magnetoelastic theory on the breathing pyrochlore Heisenberg antiferromagnet. (A--C) Magnetization curves in the strongly breathing case of $J'/J = 0.1$ for (A) $b = b' = 0.05, T/J = 0.006$, (B) $b = b' = 0.10, T/J = 0.012$, and (C) $b = b' = 0.15, T/J = 0.012$ calculated by means of the classical Monte-Carlo simulations. (D--H) Real-space magnetic structures in (D) phase I, (E) II, (F) III, and (G, H) IV, where red (blue) and yellow arrows represent spins pointing upward (downward) and canting with respect to the magnetic field $h$, respectively. In phases I$\sim$III, the canted 2-up-2-down, 3-up-1-down, and ``1-up-1-down + V"-type spin configurations are stabilized, respectively, with no translational symmetry breaking of the underlying breathing pyrochlore lattice. In phase IV, a tetrahedron-based long-range order with a six-fold magnetic unit cell size in all the three principle axes emerges. In (H), the spin configurations within three adjacent tetrahedral layers are displayed (corresponding to one fourth of the magnetic unit cell). This magnetic state is composed of the canted 2-up-2-down (purple) and 3-up-1-down tetrahedra (gray) in the ratio of 1:2.}
\label{Fig5}
\end{figure*}

\vspace{-0.2cm}
\subsection*{Classical Monte Carlo simulations}
\vspace{-0.2cm}

To get more insights on the field-induced phase transitions of LiGaCr$_{4}$O$_{8}$, we employ a microscopic magnetoelastic model on the breathing pyrochlore lattice \cite{2021_Aoy} incorporating the Einstein site phonons assuming the independent displacement of each site \cite{2006_Ber}.
The effective spin Hamiltonian
${\mathcal{H}}_{\mathrm{eff}}$ is
\begin{equation}
\label{eq:H_eff}
{\mathcal{H}}_{\mathrm{eff}} = {\mathcal{H}}_{\mathrm{0}} + {\mathcal{H}}_{\mathrm{SLC}} - h\sum_{i}S^{i},
\end{equation}
where ${\mathcal{H}}_{0}$ contains minimal exchange terms expressed as Eq.~(\ref{eq:H_0}), $h$ is the external magnetic field applied along $z$ axis, and
\begin{equation}
\begin{split}
\label{eq:H_spin-lattice coupling}
{\mathcal{H}_{\mathrm{SLC}}}&=-Jb\sum_{\langle i, j\rangle_{A}}({\mathbf S}_{i} \cdot {\mathbf S}_{j})^2-J'b'\sum_{\langle i, j\rangle_{B}}({\mathbf S}_{i} \cdot {\mathbf S}_{j})^2\\
&\quad-\sum_{i} \left\{ \frac{Jb}{4}\sum_{j\neq k\in N_{A}(i)}+\frac{J'b'}{4}\sum_{j\neq k\in N_{B}(i)} \right\} ({\mathbf S}_{i} \cdot {\mathbf S}_{j})({\mathbf S}_{i}\cdot {\mathbf S}_{k})\\
&\quad-\sqrt{JJ'bb'}\sum_{i}\sum_{j\in N_{A}(i)}\sum_{k\in N_{B}(i)}{\mathbf e}_{ij} \cdot {\mathbf e}_{ik}({\mathbf S}_{i} \cdot {\mathbf S}_{j})({\mathbf S}_{i} \cdot {\mathbf S}_{k}),\\
\end{split}
\end{equation}
where $b$ ($b'$) is a dimensionless parameter representing the strength of the spin-lattice coupling between the nearest-neighbor sites within tetrahedra {\it A} ({\it B}), $N_{A}(i)$ ($N_{B}(i)$) denotes a set of the nearest-neighbor sites of site $i$ within tetrahedra {\it A} ({\it B}), and ${\mathbf e}_{ij}$ is the unit vector oriented from site $i$ to $j$.
Possible effects from the single-ion anisotropy can be neglected here because of the quenched orbital angular momentum of the Cr$^{3+}$ ion, as indicated by the $g$-value ($g = 1.98 \sim 2.08$ \cite{2013_Oka, 2016_Lee}).
In fact, the magnetization curves of CdCr$_{2}$O$_{4}$ ($g = 2.06$ \cite{2002_Rov}), for example, for $H \parallel [100], [110]$, and [111] are almost identical \cite{2005_Chu}.
As for phonon contributions, the displacement energies of atoms are quadratic in the Hamiltonian so that they can be exactly traced out through the standard Gaussian integration (for the derivation process of Eq.~(\ref{eq:H_spin-lattice coupling}), see Supplementary Note~9).
Consequently, the phonon-mediated spin interactions ${\mathcal{H}}_{\rm SLC}$ consist of biquadratic terms favoring collinear spin configurations (first and second terms) and three-body quartic terms responsible for lifting the macroscopic degeneracy (third and last terms).
Note that Eq.~(\ref{eq:H_eff}) with $J'/J$ = 1 can reproduce a half-magnetization plateau with a 16-sublattice 3-up-1-down magnetic structure observed in HgCr$_{2}$O$_{4}$ \cite{2007_Mat} and CdCr$_{2}$O$_{4}$ \cite{2010_Mat}, where up-up-up-down chains run along all the six equivalent [110] directions \cite{2006_Ber}.
The magnetic phase diagrams for the moderately breathing case of $J'/J = 0.6$ and 0.2 are reported in Ref.~\cite{2021_Aoy}.

As mentioned above, LiGaCr$_{4}$O$_{8}$ is characterized by a large breathing anisotropy of $J'/J = 0.04$ just above $T_{\rm N}$ according to the DFT calculations, although the situation would be more complex below $T_{\rm N}$ \cite{2016_Sah}.
Here, we set $J'/J = 0.1$ as a typical value for the strongly breathing case in Eq.~(\ref{eq:H_eff}) and performed classical MC simulations to calculate the magnetization curves for various sets of the spin-lattice coupling parameters $b$ and $b'$.
Figure~\ref{Fig5}A--C shows the calculation results for a system size of $N = 16L^{3}$ spins with $L = 12$ in three typical cases: (a) $b = b' = 0.05$, (b) $b = b' = 0.10$, and (c) $b = b' = 0.15$.
Regardless of $b$ and $b'$, phase I with an 8-sublattice canted 2-up-2-down state (Fig.~\ref{Fig5}D) and phase II with a 16-sublattice 3-up-1-down state (Fig.~\ref{Fig5}E) appear followed by a higher-field phase~II' (The magnetic structure of phase II' is likely to be more complicated than a simple canted 3-up-1-down state, and its identification is beyond the scope of this work).
An important difference manifests just below phase~II.
In the weak spin-lattice coupling case of $b = b' = 0.05$, the system undergoes a first-order transition from phase~I to III with a 16-sublattice ``1-up-1-down+V"-type spin correlation (Fig.~\ref{Fig5}F), which is continuously connected to phase~II.
With increasing the spin-lattice coupling, an intermediate-field phase once disappears for $b = b' = 0.10$, and then another phase, phase~IV, appears for $b = b' = 0.15$.
Remarkably, the magnetic structure of phase IV is a mixture of the canted 2-up-2-down and 3-up-1-down units of tetrahedra {\it A} in the ratio of 1:2, forming a long-range order with a $6 \times 6 \times 6$ magnetic unit cell (Fig.~\ref{Fig5}G, H).
The detail of the ordering pattern is described in Ref.~\cite{2021_Aoy}.
We confirm that, for system sizes in which $L$ is not a multiple of six, this state cannot be obtained, and instead a metastable state with higher energy appears (Supplementary Note 10).
Note that phase~IV appears also for $J'/J = 0.2$ with larger values of $b$ and $b'$ ($b = b' \gtrsim 0.20$) but does not appear for $J'/J = 0.6$ with any values of $b$ and $b'$ \cite{2021_Aoy}.

\section*{Discussion}

From the above results, we find that the theoretical $M$--$B$ curve for the strong spin-lattice coupling case of $b = b' = 0.15$ (Fig.~\ref{Fig5}C) is qualitatively compatible with the experimental $M$--$B$ curve of LiGaCr$_{4}$O$_{8}$ (Fig.~\ref{Fig3}E) in that the two-step magnetization jump as well as the relatively wide half-magnetization plateau are reproduced.
We hence believe that the experimentally observed intermediate-field phase prior to the half magnetization plateau would be a magnetic superstructure phase induced by the strong spin-lattice coupling and large breathing anisotropy.
However, the magnetization in phase~IV amounts to $m/m_{\rm s} \approx 0.4$ (Fig.~\ref{Fig5}C), which is larger than the experimental value at $B_{\rm c2}$, $m/m_{\rm s} \approx 0.3$ (Fig.~\ref{Fig3}E).
Moreover, the present site-phonon model tends to underestimate the field width of the half-magnetization plateau \cite{2006_Ber, 2021_Aoy}, as demonstrated in {\it A}Cr$_{2}$O$_{4}$ with $J'/J = 1$ \cite{2011_Kim, 2013_Miy} and also in the present case (Supplementary Note 11).
These quantitative discrepancies may arise from the missing incorporation of the macroscopic lattice deformation in the site-phonon model.
Another microscopic magnetoelastic model assuming independent bond-length changes, i.e., the bond-phonon model, can resolve this issue (Supplementary Note 11), although it can describe neither magnetic long-range orders nor additional complicated magnetic transitions found in the site-phonon model \cite{2004_Pen, 2021_Aoy}.
The development of an extended magnetoelastic Hamiltonian is necessary to fill the gap between experiment and theory.

Finally, it is worth mentioning the relationship of the present findings to the magnetism of isolated tetrahedral clusters, namely $J' = 0$.
If we take into account quantum spins and neglect the spin-lattice coupling, the application of a magnetic field induces the quantization of a total spin number per cluster, which can be regarded as a spin crossover rather than a phase transition.
As a consequence, the $M$--$B$ curve exhibits fractional magnetization plateaus at $m/m_{\rm s} = p/2n$ for spin-$n/2$ systems ($p, n$: integer values satisfying $0 < p < 2n$).
Such a magnetization behavior was indeed observed in a spin-1/2 breathing pyrochlore antiferromagnet Ba$_{3}$Yb$_{2}$Zn$_{5}$O$_{11}$ \cite{2016_Hak} and a spin-3/2 tetrahedral-cluster compound Co$_{4}$B$_{6}$O$_{13}$ \cite{2009_Hag}.
In LiGaCr$_{4}$O$_{8}$, however, no plateau-like structure is observed at around $m/m_{\rm s} = 1/6$, i.e., $M \approx 0.5$~$\mu_{\rm B}$/Cr (Fig.~\ref{Fig3}E), suggesting that a spin crossover is masked by the intertetrahedral exchange coupling $J'$.
Also, it is obvious that if we set $J' = 0$ and $b' = 0$ in the classical Heisenberg model with the spin-lattice coupling Eq.~(\ref{eq:H_eff}), no magnetic superstructure phase appears due to the absence of spin interactions within tetrahedra {\it B}.
Three-body spin interactions across tetrahedra {\it A} and {\it B} (Fig.~\ref{Fig2}A) originating from the site-dependent local phonons, corresponding to the last term in Eq.~(\ref{eq:H_spin-lattice coupling}), play a crucial role in bringing about rich field-induced phases in the breathing pyrochlore system.

In summary, we observe a two-step magnetostructural transition between 150~T and 200~T prior to a robust half-magnetization plateau in the breathing pyrochlore antiferromagnet LiGaCr$_{4}$O$_{8}$.
Considering the magnetoelastic theory incorporating local phonon modes, the intermediate-field phase can be assigned to a spin-lattice-coupled superstructure with a three-dimensional periodic array of 3-up-1-down and canted 2-up-2-down spin molecules, which we attribute to the strong spin-lattice coupling and large breathing anisotropy.
The present work, combining the exotic experimental observations with the microscopic magnetoelastic theory in the complicated three-dimensional frustrated magnet, paves the way for further verifications of intriguing physical phenomena originating from the spin-lattice coupling and/or breathing anisotropy, both of which can be relevant in magnetic materials regardless of the geometry of the underlying crystalline lattice.
For instance, the formation of a $3 \times 3$ spin-lattice-coupled superstructure phase associated with a 1/9-magnetization plateau as well as a $\sqrt{3} \times \sqrt{3}$ one associated with a 1/3-magnetization plateau is theoretically predicted in a kagome-lattice Heisenberg antiferromagnet \cite{2022_Gen}; its experimental verification as well as theoretical studies on the effect of introducing the breathing anisotropy remain open questions.
In addition, we demonstrate that the electromagnetic induction and fiber-Bragg-grating methods are powerful tools to detect magnetic and structural phase transitions, respectively, in ultrahigh magnetic fields well above 100~T.
These techniques would be applicable to a broad range of materials such as frustrated magnets, spin-crossover systems, heavy-fermion compounds, and superconductors, leading to further flourishing of high magnetic-field science.

\section*{Methods}

\vspace{-0.2cm}
\subsection*{Sample preparation}
\vspace{-0.2cm}

Polycrystalline samples of LiGaCr$_{4}$O$_{8}$ were obtained from the same batch used in Ref.~\cite{2013_Oka}, prepared by the conventional solid state reaction method.
The powder samples were formed into rod shape with $\sim$0.8~mm in a diameter and $\sim$1.5~mm in a length using epoxy (Stycast 1266) for the magnetostriction measurement.

\vspace{-0.2cm}
\subsection*{Powder X-ray diffraction measurement and structural analysis}
\vspace{-0.2cm}

We performed the powder X-ray diffraction measurement on LiGaCr$_{4}$O$_{8}$ at 20~K using a commercial X-ray diffractometer (SmartLab, Rigaku).
The incident X-ray beam was monochromated by a Johansson-type monochromator with a Ge(111) crystal to select only Cu-$K \alpha$1 radiation.
Only a tiny amount of impurity phases were found, ensuring that the sample is of high quality.
The Rietveld analysis was performed using JANA2006 program \cite{2014_Pet}, confirming the cubic $F{\overline 4}3m$ structure.
Detailed results are shown in Supplementary Note~1.

\vspace{-0.2cm}
\subsection*{Density functional theory calculations}
\vspace{-0.2cm}

Exchange parameters of LiGaCr$_{4}$O$_{8}$ were estimated by the density-functional-theory (DFT) based energy mapping \cite{2013_Jes, 2019_Jes}.
We performed all electron DFT calculations using the full potential local orbital (FPLO) code \cite{1999_Koe}.
Note that this technique has proven to be very reliable for the breathing pyrochlore chromium spinels \cite{2020_Gen, 2019_Gho}.
We used the generalized gradient approximation (GGA) exchange and correlation functional \cite{1996_Per}.
For the electronic structure calculations, we used the $T = 20$~K crystal structure with the $F{\overline 4}3m$ space group determined in this work.
We accounted for strong correlations on the Cr $3d$ orbitals by applying a GGA+$U$ exchange correlation functional \cite{1995_Lie} for several different values of $U$ and $J_{H} = 0.72$~eV fixed following Ref.~\cite{1996_Miz}.
The fitting was performed to the Heisenberg Hamiltonian in the form ${{\mathcal{H}}}=\sum_{i<j}{\mathbf S}_{i} \cdot {\mathbf S}_{j}$, where total moments are exact multiples of 3~$\mu_{\rm B}$ as all Cr moments are exactly $S = 3/2$.
All fits are excellent.
The Weiss temperature is calculated according to $\Theta_{\rm W} = -1/3S(S+1)(3J+3J'+12J_{2} +6J_{3a} +6J_{3b})$.

\vspace{-0.2cm}
\subsection*{Magnetization and magnetostriction measurements}
\vspace{-0.2cm}

The magnetization up to 7~T was measured using a SQUID magnetometer (MPMS; Quantum Design).
The magnetization up to 51~T, 145~T, and 200~T was measured by the induction method in a non-destructive pulsed magnet, a horizontal single-turn coil (STC) system \cite{2003_Miu}, and an electromagnetic flux compression (EMFC) system \cite{2018_Nak}, respectively. 
The details of experimental setup, data, and analysis method for the magnetization measurements using the EMFC system are presented in Supplementary Notes 3--5, which include additional references \cite{1988_Tak, 1989_Ama, 1992_Got, 2003_Kir, 2017_Oka}.
The longitudinal magnetostriction up to 185 T and $\sim$600~T was measured by the optical fiber-Bragg-grating (FBG) method in the horizontal STC and EMFC systems, respectively.
Here, a relative sample-length change $\Delta L/L$ was detected by the optical filter method with the resolution of $10^{-5}$--$10^{-4}$ \cite{2017_Ike, 2023_Ike}.
The details of experimental setup, data, and analysis method for the magnetostriction measurements using the EMFC system are presented in Supplementary Notes 6--8.
All of the experiments were performed at the Institute for Solid State Physics, University of Tokyo, Kashiwa, Japan.

\vspace{-0.2cm}
\subsection*{Classical Monte Carlo simulations}
\vspace{-0.2cm}

To identify various magnetic phases appearing in the spin Hamiltonian Eq.~(\ref{eq:H_eff}), we performed classical Monte Carlo (MC) simulations in which a spin vector at each lattice site is updated in conventional random and successive over-relaxation-like processes.
In the former (latter) process, we tried to rotate a spin in a randomly proposed direction (by the angle $\pi$ around the local mean field) by using the Metropolis algorithm.
In our simulation, $2 \times 10^{6}$ MC site-sweeps were carried out at each temperature and magnetic field under the periodic boundary condition, and the first half was discarded for thermalization.
Observations were done in every 10 MC sweeps and the statistical average was taken over 4 independent runs.
We often encountered various metastable states due to very complicated interactions in the spin-lattice coupling term Eq.~(\ref{eq:H_spin-lattice coupling}), and low-temperature spin states obtained in the field-cooling, field-increasing, and field-decreasing processes were sometimes different.
In such a situation, we compare the thermal averaged values of the energy of these states and regard the lowest-energy state as the equilibrium state.
Since our cubic unit cell contains 16 sites, the total number of spins $N$ is related to the linear system size $L$ via $N = 16L^{3}$.
We have checked that the results for $L = 6$ and 12 are essentially the same and consistent with those for smaller $L$'s, e.g., $L = 1$, where the thermalization is much easier (the spin configurations for phases I, II, and III shown in Fig.~\ref{Fig5} can be described even in the small-size system of $L = 1$), so that only the result for $L = 12$ is shown in Fig.~\ref{Fig5}A--C.

\section*{Data availability}
The datasets generated during and/or analyzed during the current study are available from the corresponding author upon reasonable request.

\section*{Acknowledgements}
The authors appreciate Mr.~H. Sawabe for technical support in performing experiments using the EMFC system.
This work was partly supported by the JSPS KAKENHI Grants-In-Aid for Scientific Research (No. 20J10988 and No. 21K03469).
M.G. was a postdoctoral research fellow of the JSPS.

\section*{Author Contributions}
M.G. and Y.K. conceived and organized the project.
Y.O. synthesized polycrystalline samples of LiGaCr$_{4}$O$_{8}$.
T.Y. performed the powder XRD measurement and the Rietveld analysis.
H.O.J. performed the DFT calculations.
M.G. performed the magnetization measurement using MPMS.
H.I. performed the magnetization measurement in the non-destructive pulsed magnet under supervision of M.G. and K.K.;
M.G., D.N., and S.T. designed a magnetization probe applicable to the STC and EMFC systems.
M.G. performed the magnetization and magnetostriction measurements in the STC system under supervision of Y.H.M. and Y.K.;
M.G., A.I., Y.I., X.Z., and Y.H.M. performed the magnetization and magnetostriction measurements in the EMFC system.
K.A. performed classical MC simulations.
M.G. analyzed the experimental data and wrote the manuscript with input comments from all co-authors.

\section*{Competing interest statement}
The authors declare no conflict of interest.

\section*{Additional Information}
Supplementary Information accompanies this article.


\begin{thebibliography}{99}
\bibitem{1959_Sch} R. E. Schlier and H. E. Farnsworth, Structure and Adsorption Characteristics of Clean Surfaces of Germanium and Silicon, J. Chem. Phys. {\bf 30}, 917 (1959).
\bibitem{2006_Tok} Y. Tokura, Critical features of colossal magnetoresistive manganites, Rep. Prog. Phys. {\bf 69}, 797 (2006).
\bibitem{1975_Wil} J. A. Wilson, F. J. Di Salvo, and S. Mahajan, Charge density waves and superlattices in the metallic layered transition metal dichalcogenides. Adv. Phys. {\bf 24}, 117 (1975).
\bibitem{2021_Li} H. Li, T. T. Zhang, T. Yilmaz, Y. Y. Pai, C. E. Marvinney, A. Said, Q. W. Yin, C. S. Gong, Z. J. Tu, E. Vescovo, C. S. Nelson, R. G. Moore, S. Murakami, H. C. Lei, H. N. Lee, B. J. Lawrie, and H. Miao, Observation of Unconventional Charge Density Wave without Acoustic Phonon Anomaly in Kagome Superconductors {\it A}V$_{3}$Sb$_{5}$ ({\it A} = Rb, Cs), Phys. Rev. X {\bf 11}, 031050 (2021).
\bibitem{2002_Sch} J. Schulenburg, A. Honecker, J. Schnack, J. Richter, and H.-J. Schmidt, Macroscopic Magnetization Jumps due to Independent Magnons in Frustrated Quantum Spin Lattices, Phys. Rev. Lett. {\bf 88}, 167207 (2002).
\bibitem{2002_Kod} K. Kodama, M. Takigawa, M. Horvati\'{c}, C. Berthier, H. Kageyama, Y. Ueda, S. Miyahara, F. Becca, and F. Mila, Magnetic Superstructure in the Two-Dimensional Quantum Antiferromagnet SrCu$_{2}$(BO$_{3}$)$_{2}$, Science {\bf 298}, 395 (2002).
\bibitem{2013_Nis} S. Nishimoto, N. Shibata, and C. Hotta, Controlling frustrated liquids and solids with an applied field in a kagome Heisenberg antiferromagnet, Nat. Commun. {\bf 4}, 2287 (2013).
\bibitem{2019_Oku} R. Okuma, D. Nakamura, T. Okubo, A. Miyake, A. Matsuo, K. Kindo, M. Tokunaga, N. Kawashima, S. Takeyama, and Z. Hiroi, A series of magnon crystals appearing under ultrahigh magnetic fields in a kagom\'{e} antiferromagnet, Nat. commun. {\bf 10}, 1229 (2019).
\bibitem{1999_Kag} H. Kageyama, K. Yoshimura, R. Stern, N. V. Mushnikov, K. Onizuka, M. Kato, K. Kosuge, C. P. Slichter, T. Goto, and Y. Ueda, Exact Dimer Ground State and Quantized Magnetization Plateaus in the Two-Dimensional Spin System SrCu$_{2}$(BO$_{3}$)$_{2}$, Phys. Rev. Lett. {\bf 82}, 3168 (1999).
\bibitem{2012_Jai} M. Jaime, R. Daou, S. A. Crooker, F. Weickert, A. Uchida, A. E. Feiguin, C. D. Batista, H. A. Dabkowska, and B. D. Gaulin, Magnetostriction and magnetic texture to 100.75 Tesla in frustrated SrCu$_{2}$(BO$_{3}$)$_{2}$, Proc. Natl. Acad. Sci. U.S.A. {\bf 109}, 12404 (2012).
\bibitem{2013_Mat} Y. H. Matsuda, N. Abe, S. Takeyama, H. Kageyama, P. Corboz, A. Honecker, S. R. Manmana, G. R. Foltin, K. P. Schmidt, and F. Mila, Magnetization of SrCu$_{2}$(BO$_{3}$)$_{2}$ in Ultrahigh Magnetic Fields up to 118 T, Phys. Rev. Lett. {\bf 111}, 137204 (2013).
\bibitem{1966_Che} D. B. Chesnut, Instability of a Linear Spin Array: Application to W\"{u}rster's Blue Perchlorate, J. Chem. Phys. {\bf 45}, 4677 (1966).
\bibitem{1971_Pin} P. Pincus, Instability of the uniform antierromagnetic chain, Solid State Commun. {\bf 9}, 1971 (1971).
\bibitem{1975_Bra} J. W. Bray, H. R. Hart, Jr., L. V. Interrante, I. S. Jacobs, J. S. Kasper, G. D. Watkins, and S. H. Wee, Observation of a Spin-Peierls Transition in a Heisenberg Antiferromagnetic Linear-Chain System, Phys. Rev. Lett. {\bf 35}, 744 (1975).
\bibitem{1993_Has} M. Hase, I. Terasaki, and K, Uchinokura, Observation of the Spin-Peierls Trasition in Linear Cu$^{2+}$ (Spin-$\frac{1}{2}$) Chains in an Inorganic Compound CuGeO$_{3}$, Phys. Rev. Lett. {\bf 70}, 3651 (1993).
\bibitem{2000_Lee} S.-H. Lee, C. Broholm, T. H. Kim, W. Ratcliff II, and S.-W. Cheong, Local Spin Resonance and Spin-Peierls-like Phase Transition in a Geometrically Frustrated Antiferromagnet, Phys. Rev. Lett. {\bf 84}, 3718 (2000).
\bibitem{2002_Rov} M. T. Rovers, P. P. Kyriakou, H. A. Dabkowska, G. M. Luke, M. I. Larkin, and A. T. Savici, Muon-spin-relaxation investigation of the spin dynamics of geometrically frustrated chromium spinels, Phys. Rev. B {\bf 66}, 174434 (2002).
\bibitem{2005_Chu} J.-H. Chung, M. Matsuda, S.-H. Lee, K. Kakurai, H. Ueda, T. J. Sato, H. Takagi, K.-P. Hong, and S. Park, Statics and Dynamics of Incommensurate Spin Order in a Geometrically Frustrated Antiferromagnet CdCr$_{2}$O$_{4}$, Phys. Rev. Lett. {\bf 95}, 247204 (2005).
\bibitem{2007_Mat} M. Matsuda, H. Ueda, A. Kikkawa, Y. Tanaka, K. Katsumata, Y. Narumi, T. Inami, Y. Ueda, and S.-H. Lee, Spin-lattice instbility to a fractional magnetization state in the spinel HgCr$_{2}$O$_{4}$, Nat. Phys. {\bf 3}, 397 (2007).
\bibitem{2010_Mat} M. Matsuda, K. Ohoyama, S. Yoshii, H. Nojiri, P. Frings, F. Duc, B. Vignolle, G. L. J. A. Rikken, L.-P. Regnault, S.-H. Lee, H. Ueda, and Y. Ueda, Universal Magnetic Structure of the Half-Magnetization Phase in Cr-Based Spinels, Phys. Rev. Lett. {\bf 104}, 047201 (2010).
\bibitem{2011_Kim} S. Kimura, M. Hagiwara, T. Takeuchi, H. Yamaguchi, H. Ueda, Y. Ueda, and K. Kindo, Large change in the exchange interactions of HgCr$_{2}$O$_{4}$ under very high magnetic fields, Phys. Rev. B {\bf 83}, 214401 (2011).
\bibitem{2013_Miy} A. Miyata, S. Takeyama, and H. Ueda, Magnetic superfluid state in the frustrated spinel oxide CdCr$_{2}$O$_{4}$ revealed by ultrahigh magnetic fields, Phys. Rev. B {\bf 87}, 214424 (2013). 
\bibitem{2011_Miy} A. Miyata, H. Ueda, Y. Ueda, H. Sawabe, and S. Takeyama, Magnetic Phases of a Highly Frustrated Magnet, ZnCr$_{2}$O$_{4}$, up to an Ultrahigh Magnetic Field of 600 T, Phys. Rev. Lett. {\bf 107}, 207203 (2011).
\bibitem{2019_Ros} L. Rossi, A. Bobel, S. Wiedmann, R. K\"{u}chler, Y. Motome, K. Penc, N. Shannon, H. Ueda, and B. Bryant, Negative Thermal Expansion in the Plateau State of a Magnetically Frustrated Spinel, Phys. Rev. Lett. {\bf 123}, 027205 (2019).
\bibitem{2004_Pen} K. Penc, N. Shannon, and H. Shiba, Half-Magnetization Plateau Stabilized by Structural Distortion in the Antiferromagnetic Heisenberg Model on a Pyrochlore Lattice, Phys. Rev. Lett. {\bf 93}, 197203 (2004).
\bibitem{2006_Ber} D. L. Bergman, R. Shindou, G. A. Fiete, and L. Balents, Models of degeneracy breaking in pyrochlore antiferromagnets, Phys. Rev. B {\bf 74}, 134409 (2006).
\bibitem{2008_Wan} F. Wang and A. Vishwanath, Spin Phonon Induced Collinear Order and Magnetization Plateaus in Triangular and Kagome Antiferromagnets: Applications to CuFeO$_{2}$, Phys. Rev. Lett. {\bf 100}, 077201 (2008).
\bibitem{2013_Alb} F. A. G. Albarrac\'{i}n, D. C. Cabra, H. D. Rosales, and G. L. Rossini, Spin-phonon induced magnetic order in the kagome ice, Phys. Rev. B {\bf 88}, 184421 (2013).
\bibitem{2019_Aoy} K. Aoyama and H. Kawamura, Spin ordering induced by lattice distortions in classical Heisenberg antiferromagnets on the breathing pyrochlore lattice, Phys. Rev. B {\bf 99}, 144406 (2019).
\bibitem{2020_Miy} A. Miyata, H. Suwa, T. Nomura, L. Prodan, V. Felea, Y. Skourski, J. Deisenhofer, H.-A. Krug von Nidda, O. Portugall, S. Zherlitsyn, V. Tsurkan, J. Wosnitza, and A. Loidl, Spin-lattice coupling in a ferrimagnetic spinel: Exotic H-T phase diagram of MnCr$_{2}$S$_{4}$ up to 110 T, Phys. Rev. B {\bf 101}, 054432 (2020).
\bibitem{2020_Gen} M. Gen, Y. Okamoto, M. Mori, K. Takenaka, and Y. Kohama, Magnetization process of the breathing pyrochlore magnet CuInCr$_{4}$S$_{8}$ in ultrahigh magnetic fields up to 150 T, Phys. Rev. B {\bf 101}, 054434 (2020).
\bibitem{2021_Aoy} K. Aoyama, M. Gen, and H. Kawamura, Effects of spin-lattice coupling and a magnetic field in classical Heisenberg antiferromagnets on the breathing pyrochlore lattice, Phys. Rev. B {\bf 104}, 184411 (2021).
\bibitem{2022_Gen} M. Gen and H. Suwa, Nematicity and fractional magnetization plateaus induced by spin-lattice coupling in the classical kagome-lattice Heisenberg antiferromagnet, Phys. Rev. B {\bf 105}, 174424 (2022).
\bibitem{2017_Tsu} H. Tsunetsugu, Theory of antiferromagnetic Heisenberg spins on a breathing pyrochlore lattice, Prog. Theor. Exp. Phys. {\bf 2017}, 033I01 (2017).
\bibitem{2017_Ora} J.-C. Orain, B. Bernu, P. Mendels, L. Clark, F. H. Aidoudi, P. Lightfoot, R. E. Morris, and F. Bert, Nature of the Spin Liquid Ground State in a Breathing Kagome Compound Studied
 by NMR and Series Expansion, Phys. Rev. Lett. {\bf 118}, 237203 (2017).
\bibitem{2013_Oka} Y. Okamoto, G. J. Nilsen, J. P. Attfield, and Z. Hiroi, Breathing Pyrochlore Lattice Realized in {\it A}-Site Ordered Spinel Oxides LiGaCr$_{4}$O$_{8}$ and LiInCr$_{4}$O$_{8}$, Phys. Rev. Lett. {\bf 110}, 097203 (2013).
\bibitem{2016_Lee} S. Lee, S.-H. Do, W.-J. Lee, Y. S. Choi, M. Lee, E. S. Choi, A. P. Reyes, P. L. Kuhns, A. Ozarowski, and K.-Y. Choi, Multistage symmetry breaking in the breathing pyrochlore lattice Li(Ga,In)Cr$_{4}$O$_{8}$, Phys. Rev. B {\bf 93}, 174402 (2016).
\bibitem{2016_Sah} R. Saha, F. Fauth, M. Avdeev, P. Kayser, B. J. Kennedy, and A. Sundaresan, Magnetodielectric effects in {\it A}-site cation-ordered chromate spinels Li{\it M}Cr$_{4}$O$_{8}$ ({\it M} = Ga and In), Phys. Rev. B {\bf 94}, 064420 (2016).
\bibitem{2020_Kan} T. Kanematsu, M. Mori, Y. Okamoto, T. Yajima, and K. Takenaka, Thermal Expansion and Volume Magnetostriction in Breathing Pyrochlore Magnets Li{\it A}Cr$_{4}${\it X}$_{8}$ ({\it A} = Ga, In, {\it X} = O, S), J. Phys. Soc. Jpn. {\bf 89}, 073708 (2020).
\bibitem{2021_Pok} G. Pokharel, H. S. Arachchige, T. J. Williams, A. F. May, R. S. Fishman, G. Sala, S. Calder, G. Ehlers, D. S. Parker, T. Hong, A. Wildes, D. Mandrus, J. A. M. Paddison, and A. D. Christianson, Neutron Scattering Studies of the Breathing Pyrochlore Antiferromagnet LiGaCr$_{4}$O$_{8}$, Phys. Rev. Lett. {\bf 125}, 167201 (2021).
\bibitem{2011_Mom} K. Momma and F. Izumi, VESTA 3 for three-dimensional visu- alization of crystal, volumetric and morphology data, J. Appl. Crystallogr. {\bf 44}, 1272 (2011).
\bibitem{2008_Yar} A. N. Yaresko, Electronic band structure and exchange coupling constants in {\it A}Cr$_{2}${\it X}$_{4}$ spinels ({\it A} = Zn, Cd, Hg; {\it X} = O, S, Se), Phys. Rev. B {\bf 77}, 115106 (2008).
\bibitem{2019_Gho} P. Ghosh, Y. Iqbal, T. M\"{u}ller, R. Thomale, J. Reuther, M. J. P. Gingras, and H. O. Jeschke, Breathing chromium spinel: a showcase for a variety of pyrochlore Heisenberg Hamiltonians, npj Quantum Mater. {\bf 4}, 63 (2019).
\bibitem{2018_Nak} D. Nakamura, A. Ikeda, H. Sawabe, Y. H. Matsuda, and S. Takeyama, Record indoor magnetic field of 1200 T generated by electromagnetic flux-compression, Rev. Sci. Instrum. {\bf 89}, 095106 (2018).
\bibitem{2003_Miu} N. Miura, T. Osada, and S. Takeyama, Research in Super-High Pulsed Magnetic Fields at the Megagauss Laboratory of the University of Tokyo, J. Low. Temp. Phys. {\bf 133}, 139 (2003).
\bibitem{2012_Tak} S. Takeyama, R. Sakakura, Y. H. Matsuda, A. Miyata, and M. Tokunaga, Precise Magnetization Measurements by Parallel Self-Compensated Induction Coils in a Vertical Single-Turn Coil up to 103 T, J. Phys. Soc. Jpn. {\bf 81}, 014702 (2012).
\bibitem{2019_Gen} M. Gen, D. Nakamura, Y. Okamoto, and S. Takeyama, Ultra-high magnetic field magnetic phases up to 130~T in a breathing pyrochlore antiferromagnet LiInCr$_{4}$O$_{8}$, J. Magn. Magn. Mater. {\bf 473}, 387 (2019).
\bibitem{2017_Ike} A. Ikeda, T. Nomura, Y. H. Matsuda, S. Tani, Y. Kobayashi, H. Watanabe, and K. Sato, High-speed 100 MHz strain monitor using fiber Bragg grating and optical filter for magnetostriction measurements under ultrahigh magnetic fields, Rev. Sci. Instrum. {\bf 88}, 083906 (2017).
\bibitem{2023_Ike} A. Ikeda, Y. H. Matsuda, K. Sato, Y. Ishii, H. Sawabe, D. Nakamura, S. Takeyama, and J. Nasu, Signature of spin-triplet exciton condensations in LaCoO$_{3}$ at ultrahigh magnetic fields up to 600 T, Nat. Commun. {\bf 14}, 1744 (2023).
\bibitem{2016_Hak} T. Haku, K. Kimura, Y. Matsumoto, M. Soda, M. Sera, D. Yu, R. A. Mole, T. Takeuchi, S. Nakatsuji, Y. Kono, T. Sakakibara, L.-J. Chang, and T. Masuda, Low-energy excitations and ground-state selection in the quantum breathing pyrochlore antiferromagnet Ba$_{3}$Yb$_{2}$Zn$_{5}$O$_{11}$, Phys. Rev. B {\bf 93}, 220407(R) (2016).
\bibitem{2009_Hag} H. Hagiwara, H. Sato, M. Iwaki, Y. Narumi, and K. Kindo, Quantum magnetism of perfect spin tetrahedra in Co$_{4}$B$_{6}$O$_{13}$, Phys. Rev. B {\bf 80}, 014424 (2009).
\bibitem{2014_Pet} V. Pet\v{r}\'{i}\v{c}ek, M. Du\v{s}ek, and L. Palatinus, Crystallographic computing system JANA2006: General features, Z. Kristallogr. Cryst. Mater. {\bf 229}, 345 (2014).
\bibitem{2013_Jes} H. O. Jeschke, F. Salvat-Pujol, and R. Valent\'{i}, First-principles determination of Heisenberg Hamiltonian parameters for the spin-1/2 kagome antiferromagnet ZnCu$_{3}$(OH)$_{6}$Cl$_{2}$, Phys. Rev. B {\bf 88}, 075106 (2013).
\bibitem{2019_Jes} H. O. Jeschke, H. Nakano, and T. Sakai, From kagome strip to kagome lattice: Realizations of frustrated $S = 1$ antiferromagnets in Ti(III) fluorides, Phys. Rev. B {\bf 99}, 140410(R) (2019).
\bibitem{1999_Koe} K. Koepernik and H. Eschrig, Full-potential nonorthogonal local-orbital minimum-basis band-structure scheme, Phys. Rev. B {\bf 59}, 1743 (1999).
\bibitem{1996_Per} J. P. Perdew, K. Burke, and M. Ernzerhof, Generalized gradient approximation made simple, Phys. Rev. Lett. {\bf 77}, 3865 (1996).
\bibitem{1995_Lie} A. I. Liechtenstein, V. I. Anisimov, and J. Zaanen, Density-functional theory and strong interactions: Orbital ordering in Mott-Hubbard insulators, Phys. Rev. B {\bf 52}, R5467 (1995).
\bibitem{1996_Miz} T. Mizokawa and A. Fujimori, Electronic structure and orbital ordering in perovskite-type $3d$ transition-metal oxides studied by Hartree-Fock band-structure calculations, Phys. Rev. B {\bf 54}, 5368 (1996).
\bibitem{1988_Tak} S. Takeyama, K. Amaya, T. Nakagawa, M. Ishizuka, T. Sakakibara, T. Goto, N. Miura, Y. Ajiro, and H. Kikuchi, Magnetisation measurements in ultra-high magnetic fields produced by a single-turn coil system, J. Phys. E {\bf 21}, 1025 (1988).
\bibitem{1989_Ama} K. Amaya, S. Takeyama, T. Nakagawa, M. Ishizuka, K. Nakao, T. Sakakibara, T. Goto, N. Miura, Y. Ajiro, and H. Kikuchi, Magnetization measurements in very high plused fields produced by a single-turn coil system, Physica B {\bf 155}, 396 (1989).
\bibitem{1992_Got} T. Goto, H. Aruga Katori, T. Sakakibara, and M. Yamaguchi, Successive phase transitions in ferromagnetic YCo$_{3}$, Physica B {\bf 177}, 255 (1992).
\bibitem{2003_Kir} A. Kirste, M. Goiran, M. Respaud, J. Vanaken, J. M. Broto, H. Rakoto, M. von Ortenberg, and J. L. Garcia-Munoz, High magnetic field study of charge melting in Bi$_{1/2}$(Sr,Ca)$_{1/2}$MnO$_{3}$ perovskites: Unconventional behavior of bismuth charge ordered compounds, Phys. Rev. B {\bf 67}, 134413 (2003).
\bibitem{2017_Oka} Y. Okamoto, D. Nakamura, A. Miyake, S. Takeyama, M. Tokunaga, A. Matsuo, K. Kindo, and Z. Hiroi, Magnetic transitions under ultrahigh magnetic fields of up to 130 T in the breathing pyrochlore antiferromagnet LiInCr$_{4}$O$_{8}$, Phys. Rev. B {\bf 95}, 134438 (2017).
\end{thebibliography}
\end{document}


\title{Supplementary Information for\\``Signatures of a magnetic superstructure phase induced by ultrahigh magnetic fields in a breathing pyrochlore antiferromagnet''}

\author{M.~Gen}
\email{masaki.gen@riken.jp}
\affiliation{Institute for Solid State Physics, University of Tokyo, Kashiwa, Chiba 277-8581, Japan}
\affiliation{RIKEN Center for Emergent Matter Science (CEMS), Wako, Saitama 351-0198, Japan}

\author{A.~Ikeda}
\affiliation{Institute for Solid State Physics, University of Tokyo, Kashiwa 277-8581, Japan}
\affiliation{Department of Engineering Science, University of Electro-Communications, Chofu, Tokyo 182-8585, Japan}

\author{K.~Aoyama}
\affiliation{Department of Earth and Space Science, Graduate School of Science, Osaka University, Osaka 560-0043, Japan}

\author{H.~O.~Jeschke}
\affiliation{Research Institute for Interdisciplinary Science, Okayama University, Okayama 700-8530, Japan}

\author{Y.~Ishii}
\affiliation{Institute for Solid State Physics, University of Tokyo, Kashiwa, Chiba 277-8581, Japan}

\author{H.~Ishikawa}
\affiliation{Institute for Solid State Physics, University of Tokyo, Kashiwa, Chiba 277-8581, Japan}

\author{T.~Yajima}
\affiliation{Institute for Solid State Physics, University of Tokyo, Kashiwa, Chiba 277-8581, Japan}

\author{Y.~Okamoto}
\affiliation{Institute for Solid State Physics, University of Tokyo, Kashiwa, Chiba 277-8581, Japan}

\author{X.-G. Zhou}
\affiliation{Institute for Solid State Physics, University of Tokyo, Kashiwa, Chiba 277-8581, Japan}

\author{D.~Nakamura}
\affiliation{Institute for Solid State Physics, University of Tokyo, Kashiwa, Chiba 277-8581, Japan}
\affiliation{RIKEN Center for Emergent Matter Science (CEMS), Wako, Saitama 351-0198, Japan}

\author{S.~Takeyama}
\affiliation{Institute for Solid State Physics, University of Tokyo, Kashiwa, Chiba 277-8581, Japan}

\author{K.~Kindo}
\affiliation{Institute for Solid State Physics, University of Tokyo, Kashiwa, Chiba 277-8581, Japan}

\author{Y.~H.~Matsuda}
\affiliation{Institute for Solid State Physics, University of Tokyo, Kashiwa, Chiba 277-8581, Japan}

\author{Y.~Kohama}
\email{ykohama@issp.u-tokyo.ac.jp}
\affiliation{Institute for Solid State Physics, University of Tokyo, Kashiwa, Chiba 277-8581, Japan}

\begin{abstract}

\end{abstract}

\maketitle

\onecolumngrid

\vspace{+2.0cm}
This Supplementary Information includes contents as listed below:\\
\\
Note~~~1.~Structural analysis of LiGaCr$_{4}$O$_{8}$ at 20 K\\
Note~~~2.~Magnetizatinon data up to 7~T obtained in MPMS\\
Note~~~3.~Details of magnetization measurement technique in the EMFC system\\
Note~~~4.~All the magnetization data obtained in the EMFC system\\
Note~~~5.~Analysis of the magnetization data obtained in the EMFC system\\
Note~~~6.~Details of magnetostriction measurement technique in the EMFC system\\
Note~~~7.~All the magnetostriction data obtained in the EMFC system\\
Note~~~8.~Analysis of the magnetostriction data obtained in the EMFC system\\
Note~~~9.~Site-phonon model on the breathing pyrochlore Heisenberg antiferromagnet\\
Note~10.~System size dependence of the calculated results on Monte Carlo simulations\\
Note~11.~Magnetic-field width of the half-magnetization plateau\\

\newpage
\subsection*{\label{Sec1} \large Note 1. Structural analysis of LiGaCr$_{4}$O$_{8}$ at 20 K}

\begin{figure}[h]
\centering
\includegraphics[width=0.65\linewidth]{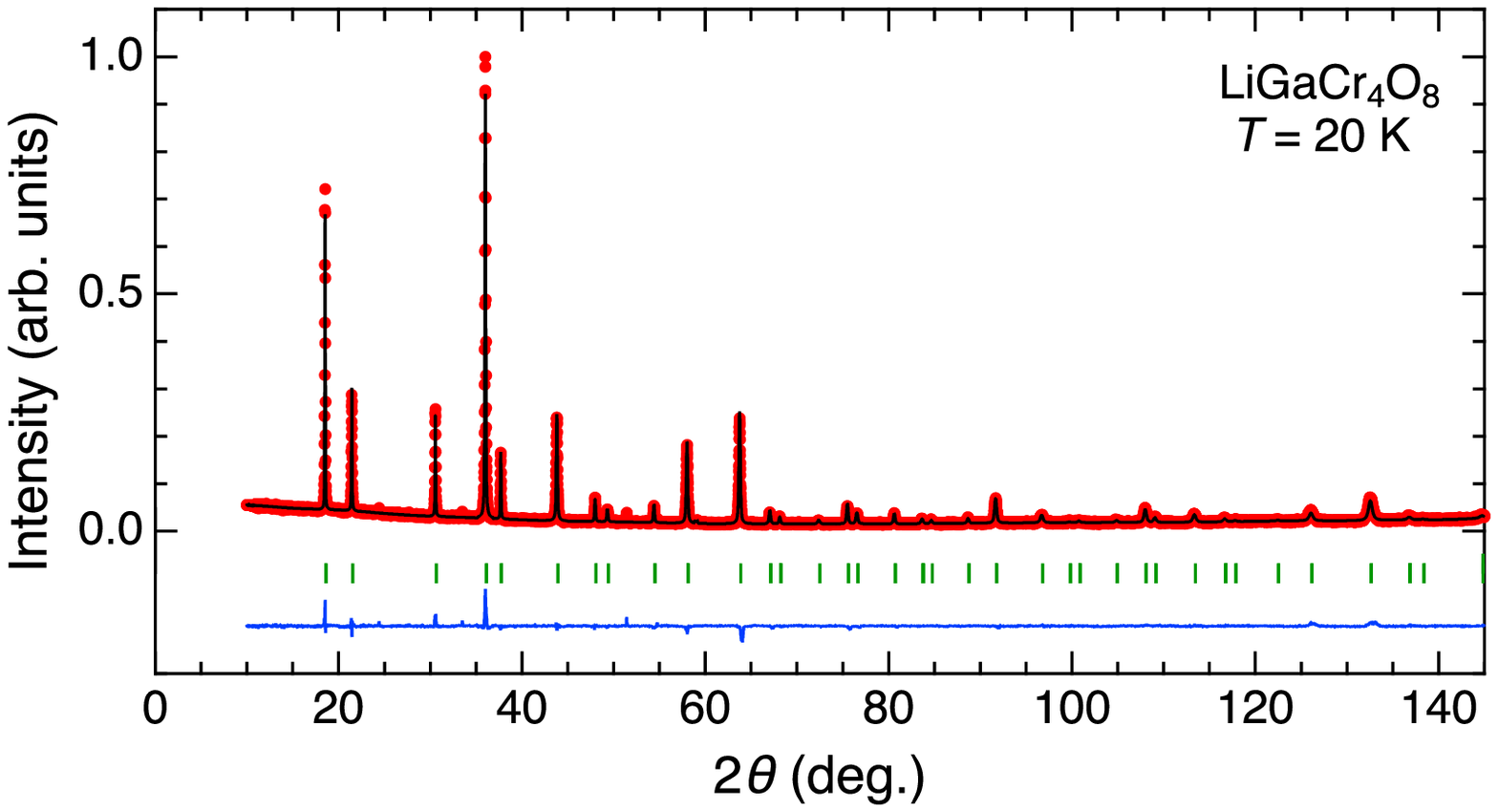}
\renewcommand{\thefigure}{S\arabic{figure}}
\caption{Refinement result of the powder XRD patterns of LiGaCr$_{4}$O$_{8}$ at 20~K. The red filled circles and the black solid line show the experimental data points and the calculated pattern, respectively. The vertical bars indicate the positions of the structure Bragg peaks. The blue line shows the difference in the experimental and calculated intensities.}
\label{FigS1}
\end{figure}

The crystallographic parameters of LiGaCr$_{4}$O$_{8}$ polycrystalline samples used in this study at 293~K was reported in Ref.~\cite{2013_Oka}.
Here, we investigated the structural change in the paramagnetic region by measuring the powder X-ray diffraction (XRD) patterns at 20~K using a commercial X-ray diffractometer (SmartLab, Rigaku).
Figure~\ref{FigS1} shows the observed powder XRD pattern and the result of the Rietveld analysis based on a structural model with the cubic space group $F{\overline 4}3m$.
The crystallographic parameters obtained by assuming the occupancy of each site to be 1 are summarized in Table~\ref{tab:S1}.

\vspace{+1.0cm}
\begin{table}[h]
\renewcommand{\arraystretch}{1.2}
\renewcommand{\thetable}{S\arabic{table}}
\caption{Crystallographic parameters of LiGaCr$_{4}$O$_{8}$ polycrystalline samples at 20~K. The lattice constant is $a = 8.24031(2)$~\AA.}
\vspace{+0.1cm}
\begin{tabular}{ccccccc} \hline\hline
\multicolumn{7}{l}{~~~~$R_{\rm p} = 5.13\%, R_{\rm wp} = 6.13\%, S = 1.260$} \\
~~~ & ~~~ & ~~~$x$~~~ & ~~~$y$~~~ & ~~~$z$~~~ & ~~Occupancy~~ & ~~~100$U_{\rm iso}$ (\AA$^{2}$)~~~ \\ \hline
~~~Li~~~ & ~~~$4a$~~~ & ~~~0~~~ & ~~~0~~~ & ~~~0~~~ & ~~~1~~~ & ~~~1~~~ \\
~~~Ga~~~ & ~~~$4d$~~~ & ~~~3/4~~~ & ~~~3/4~~~ & ~~~3/4~~~ & ~~~1~~~ & ~~~0.62(4)~~~ \\
~~~Cr~~~ & ~~~$16e$~~~ & ~~~0.37526(15)~~~ & ~~~$x$~~~ & ~~~$x$~~~ & ~~~1~~~ & ~~~0.54(3)~~~ \\
~~~O1~~~ & ~~~$16e$~~~ & ~~~0.1376(2)~~~ & ~~~$x$~~~ & ~~~$x$~~~ & ~~~1~~~ & ~~~0.78(5)~~~ \\
~~~O2~~~ & ~~~$16e$~~~ & ~~~0.6275(4)~~~ & ~~~$x$~~~ & ~~~$x$~~~ & ~~~1~~~ & ~~~0.78(5)~~~ \\ \hline\hline
\end{tabular}
\label{tab:S1}
\end{table}

\vspace{+1.5cm}

\newpage
\subsection*{\label{Sec2} \large Note 2. Magnetization data up to 7~T obtained in MPMS}

\begin{figure}[h]
\centering
\includegraphics[width=0.9\linewidth]{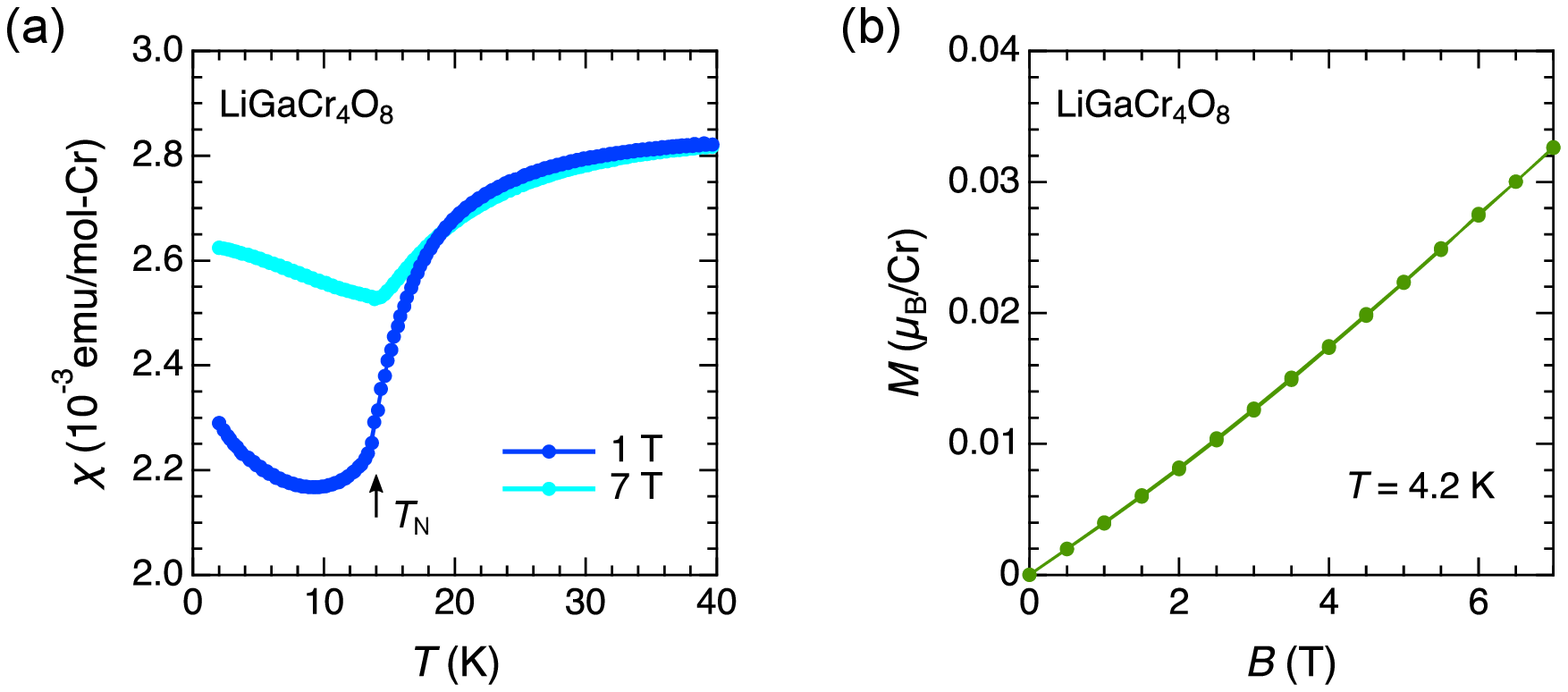}
\renewcommand{\thefigure}{S\arabic{figure}}
\caption{(a) Temperature dependence of the field-cooled magnetic susceptibility $\chi$ of LiGaCr$_{4}$O$_{8}$ measured in a magnetic field of 1~T (blue) and 7~T (cyan). (b) Magnetization curves of LiGaCr$_{4}$O$_{8}$ measured at 4.2~K up to 7~T.}
\label{FigS2}
\end{figure}

Figure~\ref{FigS2} summarizes the magnetization data of LiGaCr$_{4}$O$_{8}$ polycrystalline samples obtained using a SQUID magnetometer (MPMS; Quantum Design).
Figure~\ref{FigS2}(a) shows the temperature dependence of the magnetic susceptibility $\chi$ measured at 1~T and 7~T.
$\chi$ rapidly drops at $T_{\rm N} \approx 14$~K at 1~T, indicating an antiferromagnetic (AFM) phase transition.
Figure~\ref{FigS2}(b) shows the magnetization curve measured up to 7~T.
A slightly upturn behavior is seen at around 2~T, suggesting a weak spin-flop transition or a crossover.
Note that the coexistence of tetragonal and cubic phases are reported in the zero-field AFM state below $T_{\rm N}$ \cite{2016_Sah}.
The applied magnetic field may favor one of them, resulting in a change in the slope of the $M$--$B$ curve.

\vspace{+1.0cm}

\newpage
\subsection*{\label{Sec3} \large Note 3. Details of magnetization measurement technique in the EMFC system}

In many frustrated magnets, the energy scale of the overall AFM exchange interactions reaches several hundreds of kelvin.
Therefore, magnetization measurements in ultrahigh magnetic fields above 100~T are indispensable to understand their underlying physics.
Over the past 30 years, the electromagnetic induction method using the single-turn coil (STC) system has been improved step by step \cite{1988_Tak, 1989_Ama, 1992_Got, 2003_Kir, 2012_Tak, 2017_Oka, 2019_Gen}, and now precise magnetization measurements up to 150~T are available \cite{2020_Gen}.
However, the induction method in much higher fields using the electromagnetic flux compression (EMFC) system remains very challenging because of (i) the difficulty in the magnetic-field generation itself, (ii) the complete destruction inside the magnet coil by a single shot, and (iii) the poor reproducibility of the magnetic-field waveform.
Thus, it is impractical to experimentally subtract the uncompensated $dB/dt$ component from the $dM/dt$ waveform by the conventional way adopted in the non-destructive pulsed magnet and the STC system, i.e., performing the measurement twice with sample-in and sample-out conditions under the same magnetic-field waveform.

\begin{figure}[h]
\centering
\includegraphics[width=0.8\linewidth]{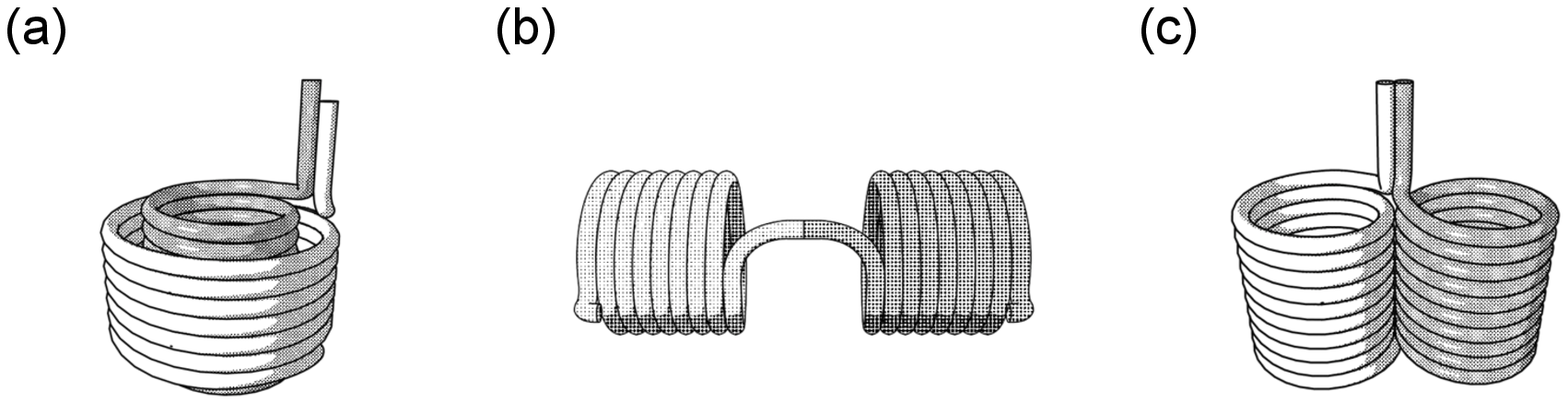}
\renewcommand{\thefigure}{S\arabic{figure}}
\caption{Schematics of self-compensated $M$ pickup coils with three different geometries: (a) coaxial-type, (b) straight-type, and (c) parallel-type.}
\label{FigS3}
\end{figure}

\begin{figure}[h]
\centering
\includegraphics[width=0.8\linewidth]{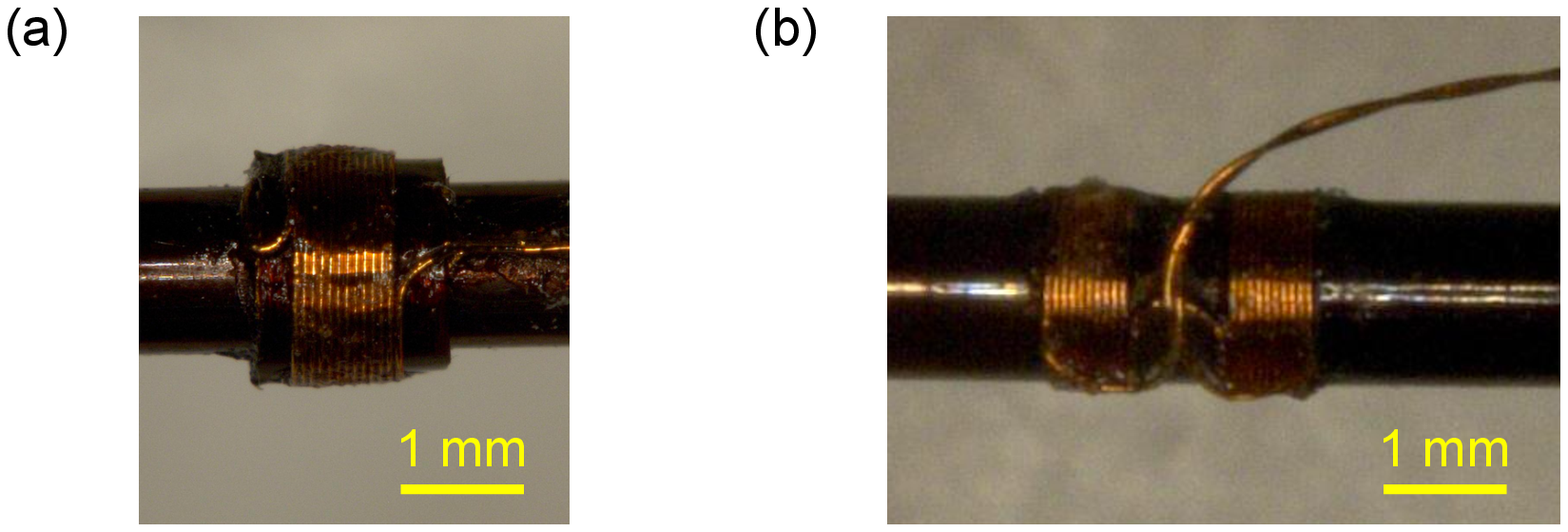}
\renewcommand{\thefigure}{S\arabic{figure}}
\caption{Photographs of self-compensated $M$ pickup coils: (a) coaxial-type, (b) straight-type.}
\label{FigS4}
\end{figure}

For detecting $dM/dt$ signal in the EMFC system, we adopted two types of handmade self-compensated $M$ pickup coils, coaxial-type [Fig.~\ref{FigS3}(a)] and straight-type [Fig.~\ref{FigS3}(b)].
These shapes have an advantage compared to the parallel-type [Fig.~\ref{FigS3}(c)] in that the special inhomogeneity of the generated magnetic field in a radial direction can be suppressed.
Indeed, it has been experimentally demonstrated that at above 100~T the field generation space is already smaller than 2~cm in a diameter, and its special inhomogeneity significantly increases in a radial direction \cite{2014_Nak, 2018_Nak}.
The $M$ pickup coils were made of a polyimide-amide enameled copper wire (AIW wire, TOTOKU Electric Co. Ltd.) with an outer diameter of 0.06~mm and a withstand voltage of $\sim$1~kV, which were wound around a long Kapton tube.
The typical designs are as follows: 21 turns (or 29 turns) for an inner coil with a diameter of 1.3~mm and 13 turns (or 18 turns) for an outer coil with a diameter of 1.7~mm for the coaxial-type, and 10 turns for both left and right coils with a diameter of 1.7~mm for the straight-type (see also the photos in Fig.~\ref{FigS4}).
It was confirmed that these $M$ pickup coils achieve a high compensation ratio of more than 99.9~\% and can survive without dielectric breakdown until they are mechanically destroyed by the implosion of a liner at around the maximum field of 400--600~T.
The magnetic field was simultaneously measured by another $B$ pickup coil wound outside the $M$ pickup coil.

\vspace{+0.5cm}

\vspace{+0.5cm}

\begin{figure}[h]
\centering
\includegraphics[width=0.8\linewidth]{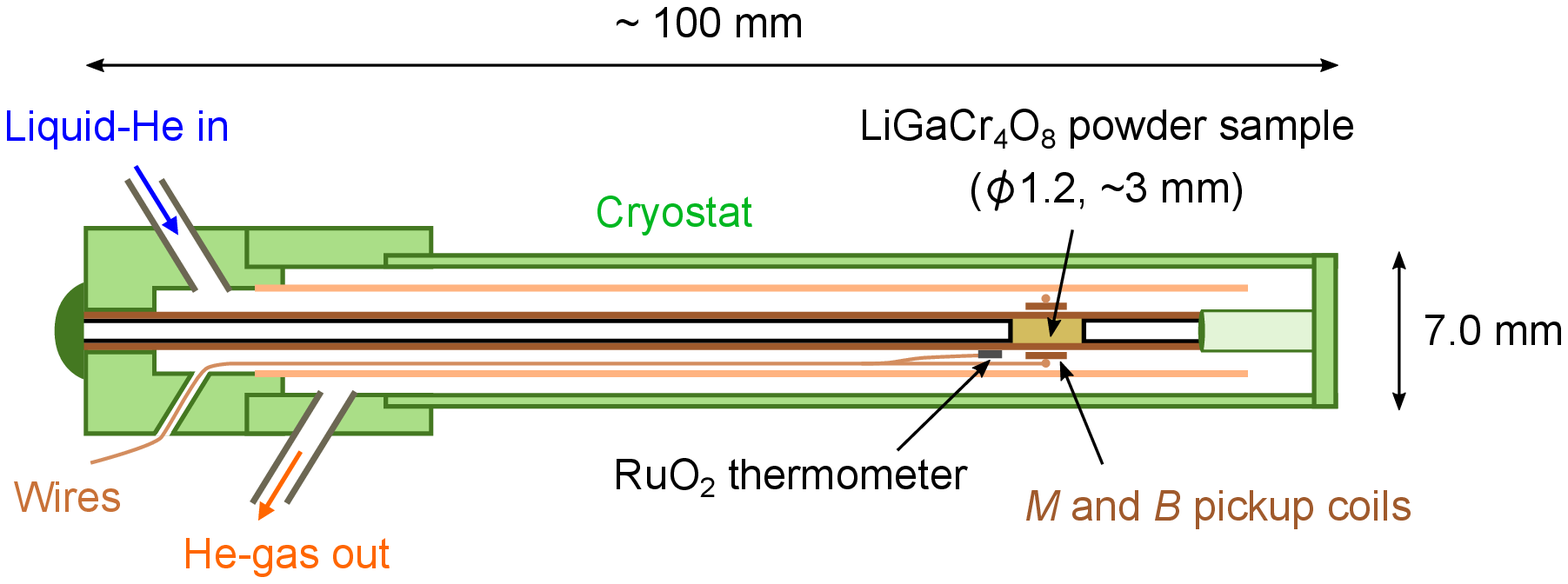}
\renewcommand{\thefigure}{S\arabic{figure}}
\caption{Cross-sectional view of the liquid-$^{4}$He flow cryostat equipped with the $M$ and $B$ pickup coils, RuO$_{2}$ thermometer, and sample.}
\label{FigS5}
\end{figure}

\newpage
In order to cool the sample down to $\sim$5~K, the $M$ and $B$ pickup coils were installed in a double-layerd liquid-$^{4}$He flow cryostat made of glass-epoxy (G10), as illustrated in Fig.~\ref{FigS5}.
The LiGaCr$_{4}$O$_{8}$ powder samples were packed inside the Kapton tube wound with the $M$ pickup coil in a length of 3~mm and was fixed between sticks made of G10 from both sides.
To monitor the sample temperature, a RuO$_{2}$ tip thermometer was glued to the Kapton tube in the vicinity of the sample.

\vspace{+1.0cm}

\newpage
\subsection*{\label{Sec4} \large Note 4. All the magnetization data obtained in the EMFC system}

\begin{figure}[h]
\centering
\includegraphics[width=\linewidth]{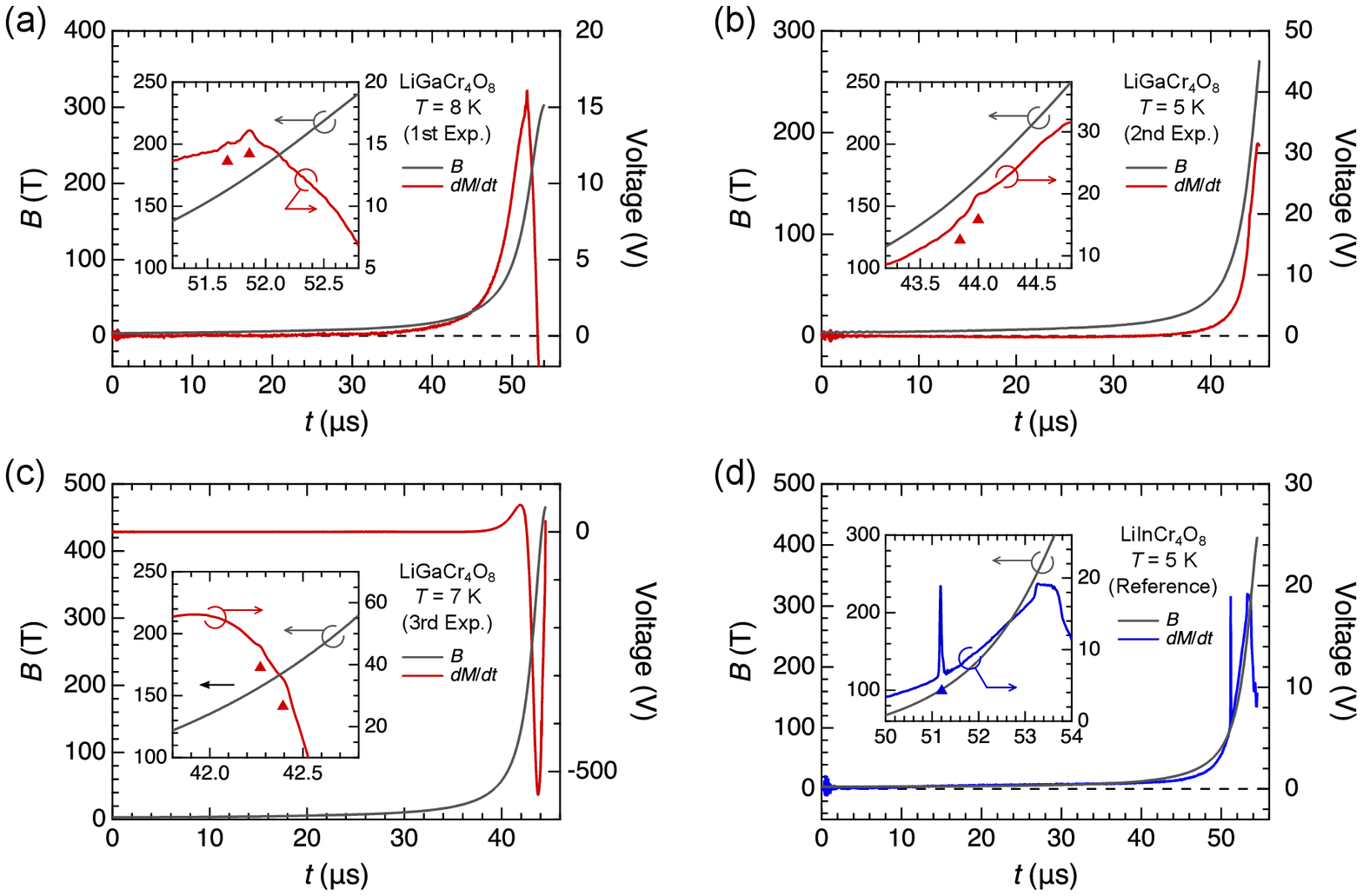}
\renewcommand{\thefigure}{S\arabic{figure}}
\caption{Temporal evolutions of the magnetic field (gray) and the induction voltage detected by the $M$ pickup coil (red or blue) obtained in the EMFC system. The measurements were performed for LiGaCr$_{4}$O$_{8}$ in (a)--(c) and for LiInCr$_{4}$O$_{8}$ in (d). In all the panels, the inset displays an enlarged view around phase transitions, as denoted by upward triangles in $dM/dt$ data.}
\label{FigS6}
\end{figure}

We performed the magnetization measurement on LiGaCr$_{4}$O$_{8}$ using the EMFC system in total three times.
Here, the coaxial-type $M$ pickup coil was employed for the 1st and 2nd experiments, while the straight-type for the 3rd.
The temporal evolutions of the magnetic field (gray) and the induction voltage detected by the $M$ pickup coil (red) for each experiment are shown in Figs.~\ref{FigS6}(a)--\ref{FigS6}(c), where the measurement temperature was $\sim$8~K, $\sim$5~K, and $\sim$7~K, respectively.
In the main text, we show the data of the 2nd experiment as representative (Figs.~3C and 3E).
As denoted by triangles in the insets, the observation of double-hump $dM/dt$ anomalies around 160--170~T are reproduced in all the experiments, indicating that the two-step transition is intrinsic.
We note that the background component is much larger and its turnaround behavior occurs at a lower magnetic field in the $dM/dt$ waveform for the 3rd experiment, making the intrinsic $dM/dt$ anomalies blurred.
Thus, we conclude that the coaxial-type $M$ pickup coil is superior to the straight-type for detecting the intrinsic $M$ signal sensitively up to as high magnetic fields as possible.

In addition, we also performed the magnetization measurement on LiInCr$_{4}$O$_{8}$ as a reference sample, which is known to exhibit a first-order phase transition to the 1/2-magnetization plateau phase at 100 T \cite{2017_Oka, 2019_Gen}, in order to quantitatively estimate the absolute value of the magnetization jump for LiGaCr$_{4}$O$_{8}$ as discussed in the next section.
The coaxial-type $M$ pickup coil was employed in this measurement.
The temporal evolutions of the magnetic field (gray) and the induction voltage detected by the $M$ pickup coil (blue) at 5~K are shown in Fig.~\ref{FigS3}(d).
As denoted by a triangle in the inset, a sharp $dM/dt$ peak is observed at 100~T, which is consistent with the previous observations obtained in the STC system \cite{2017_Oka, 2019_Gen}.
Here, additional $dM/dt$ anomalies are observed in a high-field region around 250~T (at 53.2~$\mu$s), presumably indicating a phase transition.
The interpretation of these observations is out of the scope of this paper.

\vspace{+0.5cm}

\newpage
\subsection*{\label{Sec5} \large Note 5. Analysis of the magnetization data obtained in the EMFC system}

We here introduce the procedure to obtain the $M$--$B$ curve of LiGaCr$_{4}$O$_{8}$ (Fig.~3E) from the raw data (Fig.~3C), which is equivalent to Fig.~\ref{FigS6}(b).\\

\begin{figure}[h]
\centering
\includegraphics[width=0.95\linewidth]{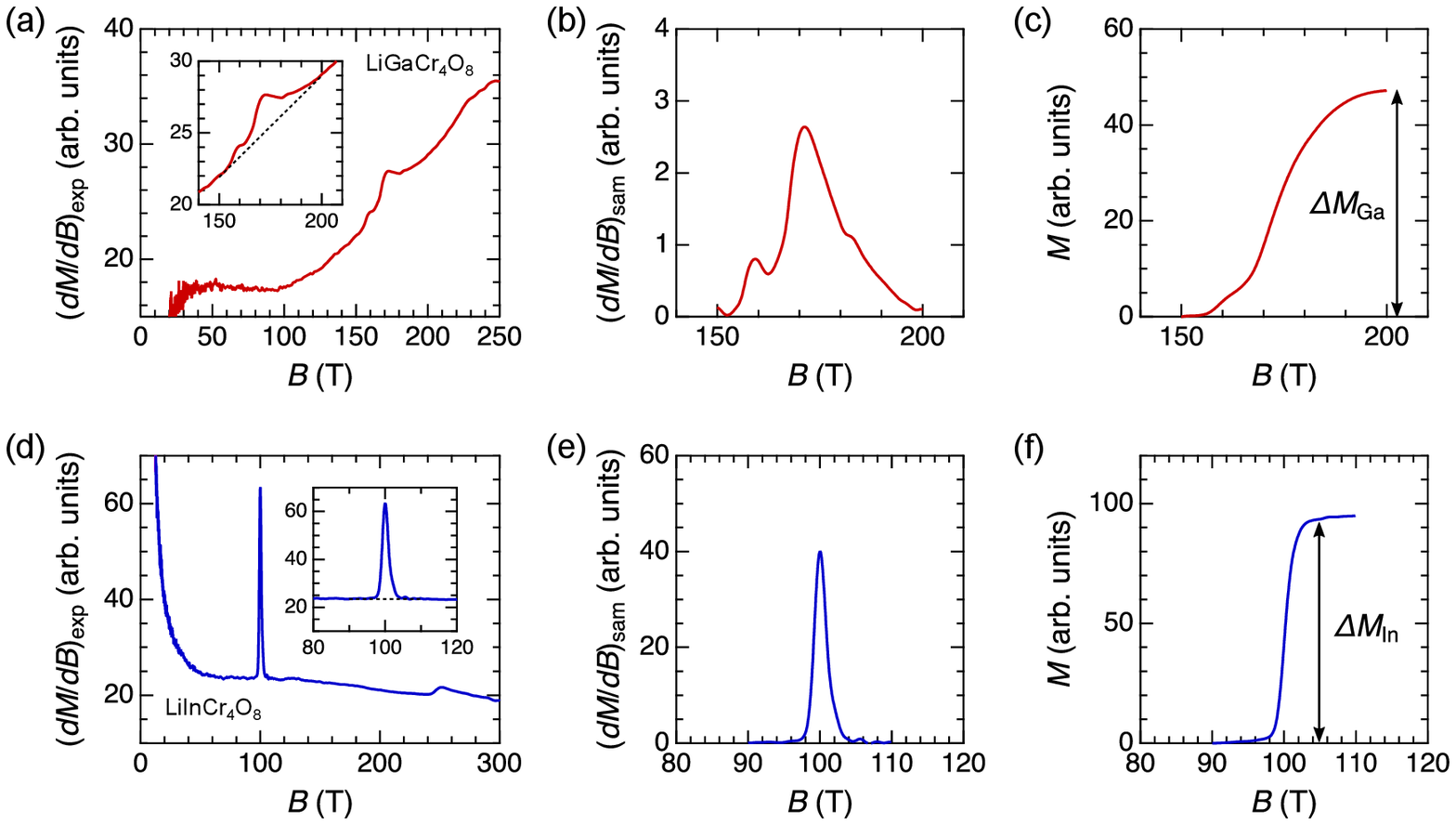}
\renewcommand{\thefigure}{S\arabic{figure}}
\caption{Procedure for analyzing the magnetization data obtained in the EMFC system for [(a)--(c)] LiGaCr$_{4}$O$_{8}$ and [(d)--(f)] LiInCr$_{4}$O$_{8}$. See the text for details.}
\label{FigS7}
\end{figure}

\hspace{-0.5cm}
[Step 1] First, we derive the experimental $dM/dB$ waveform [$(dM/dB)_{\rm exp}$] for LiGaCr$_{4}$O$_{8}$ and LiInCr$_{4}$O$_{8}$ by dividing $dM/dt$ measured by the $M$ pickup coil by $dB/dt$ measured by the $B$ pickup coil.
The obtained $(dM/dB)_{\rm exp}$ waveforms for LiGaCr$_{4}$O$_{8}$ and LiInCr$_{4}$O$_{8}$ are shown in Figs.~\ref{FigS7}(a) and \ref{FigS7}(d), respectively.\\

\hspace{-0.5cm}
[Step 2] We approximate a linear background in the $(dM/dB)_{\rm exp}$ waveform in the field region of 150--200~T for LiGaCr$_{4}$O$_{8}$ and 90--110 T for LiInCr$_{4}$O$_{8}$ as shown by the black dashed line in the insets of Figs.~\ref{FigS7}(a) and \ref{FigS7}(d), respectively.
Then, we obtain quasi-intrinsic $dM/dB$ waveforms [$(dM/dB)_{\rm sam}$] by subtracting the linear background from the $(dM/dB)_{\rm exp}$ waveforms [Figs.~\ref{FigS7}(b) and \ref{FigS7}(e)].\\

\hspace{-0.5cm}
[Step 3] By integrating $(dM/dB)_{\rm sam}$ as a function of $B$, $M$--$B$ curves are obtained as shown in Figs.~\ref{FigS7}(c) and \ref{FigS7}(f).
Here, $\Delta M_{\rm Ga}$ and $\Delta M_{\rm In}$ represent the magnitudes of a magnetization jump accompanied by a phase transition to the 1/2-magnetization plateau.
The vertical axes are shown by the same units to enable the quantitative comparison of the experimentally-detected increases in the total magnetic moments.
For LiInCr$_{4}$O$_{8}$, the magnitude of the magnetization jump at 100~T is estimated to be $0.6 \pm 0.1$~$\mu_{\rm B}$/Cr by the previous magnetization measurements in the STC system \cite{2017_Oka, 2019_Gen}.
The corresponding change in Fig.~\ref{FigS7}(f) is $\Delta M_{\rm In} \approx 93$, while $\Delta M_{\rm Ga} \approx 47$ in Fig.~\ref{FigS7}(c) as for LiGaCr$_{4}$O$_{8}$.
Here, the difference in the number of turns between the inner and outer coils of the employed coaxial $M$ pickup coil was 8 for LiGaCr$_{4}$O$_{8}$ and 11 for LiInCr$_{4}$O$_{8}$.
Assuming that the sensitivity of the magnetization detection is proportional to the difference in the number of turns, the magnitude of the magnetization jump in the field region of 150--200~T for LiGaCr$_{4}$O$_{8}$ can be calculated as $(0.6 \pm 0.1) \times (47/93) \times (11/8) = 0.42 \pm 0.07$~$\mu_{\rm B}$/Cr.\\

\hspace{-0.5cm}
[Step 4] We extrapolate the linear $M$--$B$ curve obtained in the HSTC system as drawn by the black dashed line in Fig.~3E, which can be expressed as $M_{\rm linear}(B)~[\mu_{\rm B}/{\rm Cr}] = 0.0048 B$~[T].
The final $M$--$B$ curve in the field region of 150--200~T is derived by adding the $M$--$B$ curve obtained in Step 3 to the linear component $M_{\rm linear}(B)$.

\vspace{+0.5cm}

\newpage
\subsection*{\label{Sec6} \large Note 6. Details of magnetostriction measurement technique in the EMFC system}

\begin{figure}[h]
\centering
\includegraphics[width=\linewidth]{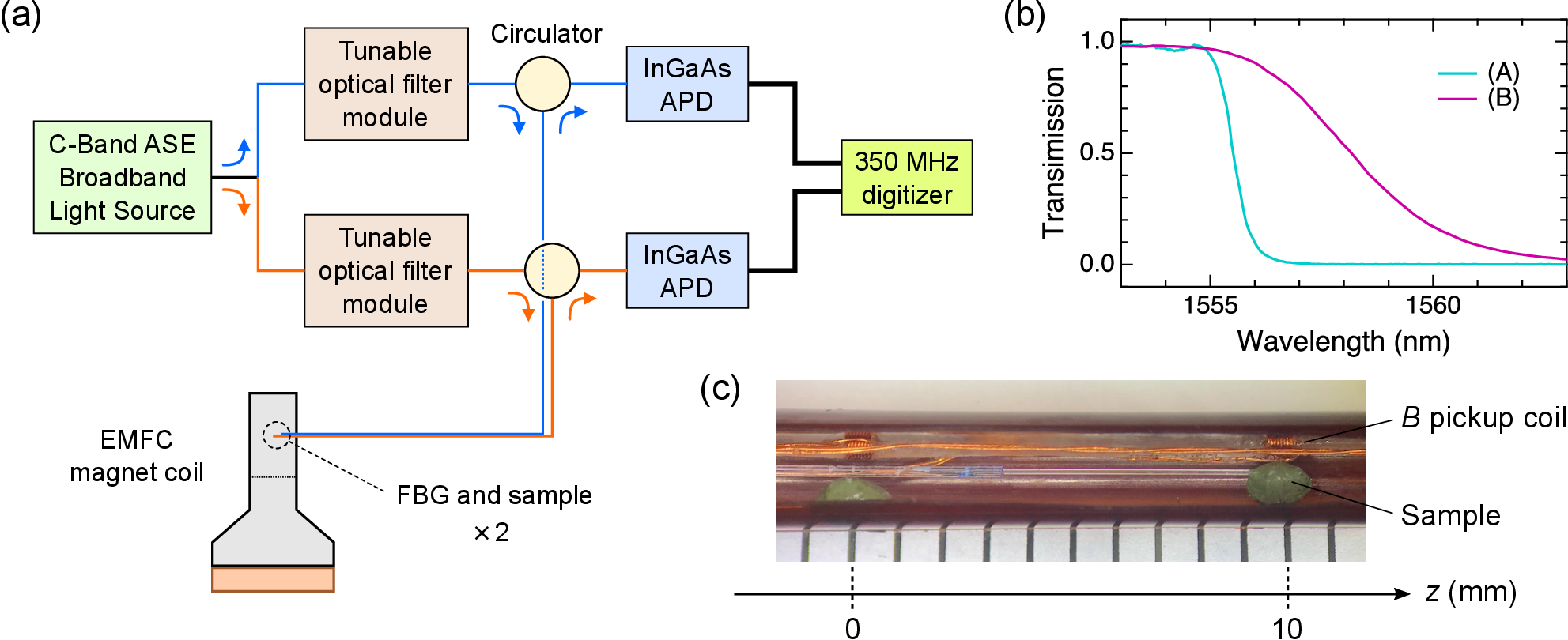}
\renewcommand{\thefigure}{S\arabic{figure}}
\caption{(a) Block diagram of the renewed high-speed strain monitoring system using the FBG in combination with the optical filter method, enabling simultaneous measurement of two channels. (b) Transmission spectra of the optical filters (A) and (B). (c) Photograph of two FBGs glued to two rod-shaped LiGaCr$_{4}$O$_{8}$ polycrystalline samples placed in a cut-open Kapton tube with $\phi$2.5. This setup was used for the experiment of Shot I.}
\label{FigS8}
\end{figure}

For the magnetostriction measurement, we employed the original high-speed strain gauge using fiber-Bragg-grating (FBG) technique and optical filter method implemented in ISSP \cite{2017_Ike, 2023_Ike}.
In order to efficiently perform the experiment, we have recently extended the FBG measurement system to enable simultaneous measurements of two channels, as schematically shown in Fig.~\ref{FigS8}(a).
Here, C-Band ASE Broadband Light Source (TLA-0016, Amonics Ltd.) is employed.
The light source is immediately separated into two branches, and two independent optical filters are applied.
The filtered light source illuminates the FBG glued to the sample after passing through an optical circulator.
The reflected light from each FBG is guided by the optical circulator and detected with independent InGaAs avalanche photodetectors (APD) (Thorlabs APD430C, conversion gain: $1.8 \times 10^{5}$~V/W, bandwidth: 400~MHz).
Finally, the data are acquired using a single oscilloscope (Lecroy HDO4034A, bandwidth: 350~MHz).
The development of multi-channel measurements is a major advance in view of the difficulty of performing repetitive experiments with the EMFC system.

In the present work, we performed the FBG experiment twice in the EMFC systems (Shots I and II).
In each shot, we placed two LiGaCr$_{4}$O$_{8}$ samples (rod shape with $\sim$0.8~mm in a diameter and $\sim$1.5~mm in a length formed using Stycast 1266) on the probe [Fig.~\ref{FigS8}(c)] and measured their magnetostriction simultaneously.
We show all the data in Supplementary Note~7 and their measurement conditions in Table~\ref{tab:S2} (shown in the next page).
Data \#3 and \#5 were obtained in Shot I, and Data \#4 and \#6 in Shot II.
We show Data \#3 $\sim$ \#5 in Fig.~4 in the main text.
The sample was located at the center position in the axial direction with respect to the magnet coil for Data \#4 and \#5, whereas the sample was 10 and 5~mm away from the center in the axial direction for Data \#3 and \#6, respectively.
We also placed two $B$ pickup coils made of AIW wire to measure the magnetic field at each sample position and one RuO$_{2}$ tip thermometer.

In the optical filter method, the shift of the Bragg wavelength is converted to the change in the intensity of the optical signal (for details, see Ref.~\cite{2017_Ike}).
Here, we used two types of optical filters, (A) 1560SPF and (B) 1551BPF (Koshin Kogaku TFM/FC), both of which allow the tuning of the cut-off wavelength by $\pm 10$~nm around $\lambda = 1555$~nm.
Transmission spectra of Filters (A) and (B) are displayed in Fig.~\ref{FigS8}(b).
The former allows high sensitivity of $\Delta L/L \sim 1 \times 10^{-5}$ and a narrow dynamic range of $\Delta L/L \sim 6 \times 10^{-4}$, whereas the latter allows low sensitivity of $\Delta L/L \sim 5 \times 10^{-5}$ and a broad dynamic range of $\Delta L/L \sim 3 \times 10^{-3}$.
We used Filter (A) for Data~\#3 and \#6 and (B) for Data~\#4 and \#5.
Note that we used Filter (A) to obtain Data~\#1 and \#2 in the STC system (Fig.~3F in the main text). 

\vspace{+1.0cm}

\newpage
\subsection*{\label{Sec7} \large Note 7. All the magnetostriction data obtained in the EMFC system}

\begin{figure}[h]
\centering
\includegraphics[width=\linewidth]{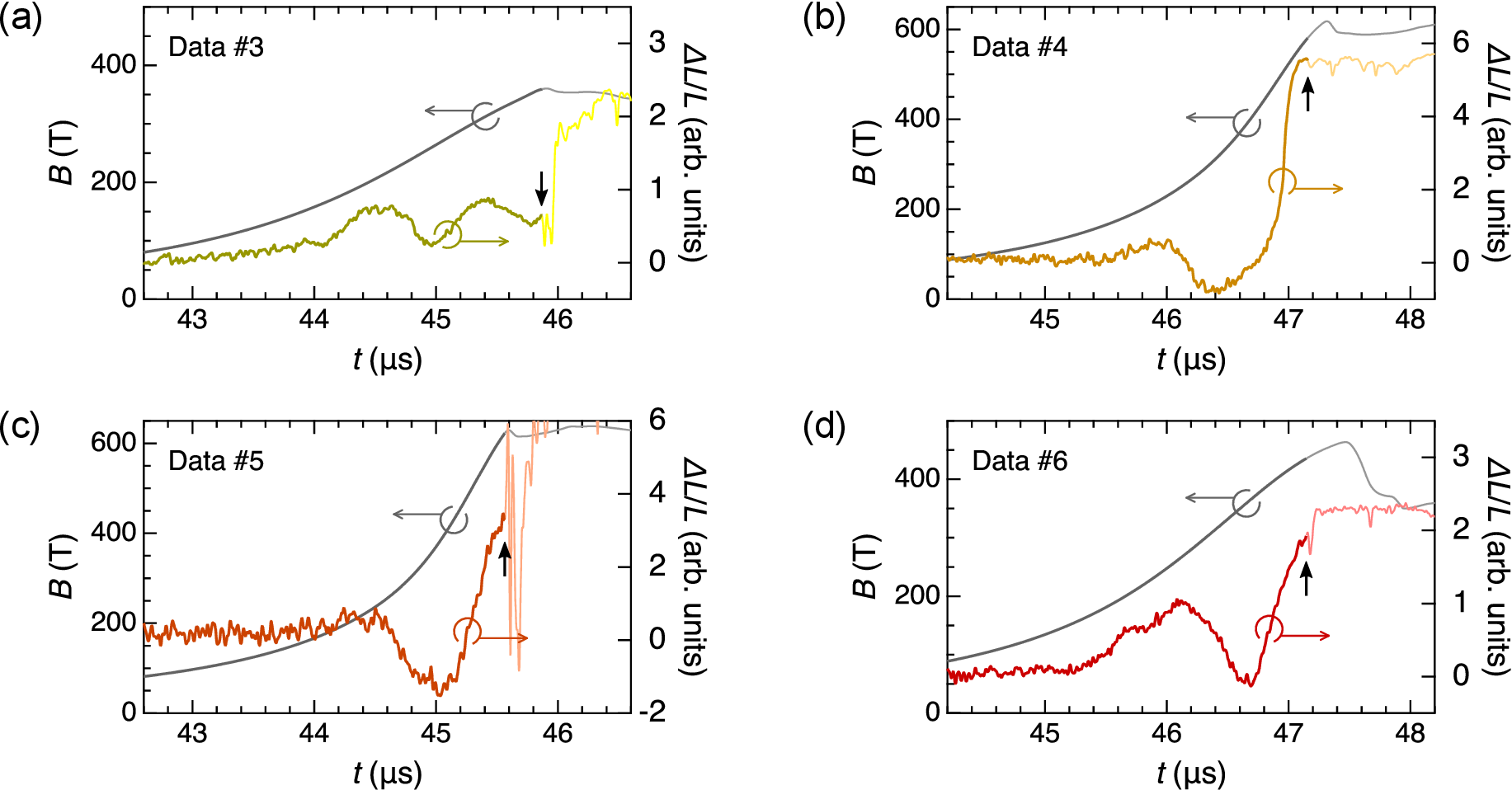}
\renewcommand{\thefigure}{S\arabic{figure}}
\caption{Temporal evolutions of the magnetic field (left axis) and the relative sample-length change $\Delta L/L$ (right axis) obtained in the EMFC system for (a) Data \#3, (b) Data \#4, (c) Data \#5, and (d) Data \#6. The measurement temperature was $\sim$5~K for all data. The thick (thin) lines correspond to the data before (after) the timing of the mechanical breakdown of the probe, which is indicated by black arrows.}
\label{FigS9}
\end{figure}

Figure~\ref{FigS9} shows the raw data of the relative sample-length change $\Delta L/L$ obtained by the FBG strain gauge as well as the time dependence of the field strength in the EMFC system.
All the measurements were performed at $\sim$5~K.
The observed $\Delta L/L$--$t$ curves are significantly affected by oscillation noise above 200~T, which should be caused by the shock wave propagating inside the sample immediately after the ultrafast magnetostructural transitions at $B_{\rm c1}$ and $B_{\rm c2}$ \cite{2017_Ike}.
For Data \#3, as the oscillating amplitude remains almost constant between 200~T and $B_{\rm max}$, no additional phase transition accompanied by an upturn behavior of $\Delta L/L$ is likely to occur.
For Data \#4 and \#5 with $B_{\rm max} \approx 600$~T, on the other hand, a $\Delta L/L$ increase much larger than the negative $\Delta L/L$ behavior caused by the mechanical vibration is reproducibly observed, signaling the occurrence of a phase transition from the 3-up-1-down to a higher-field spin-canted phase.

We note that no oscillation noise is observed in Data \#1 and \#2 obtained using the STC system, as shown in Figs.~3D and 3F in the main text.
This would be because (i) the transition to the 3-up-1-down phase occurs only halfway through due to insufficient $B_{\rm max}$, and (ii) the transition occurs near $B_{\rm max}$ where the $dB/dt$ becomes small.

\vspace{+0.5cm}
\begin{table}[h]
\renewcommand{\arraystretch}{1.2}
\renewcommand{\thetable}{S\arabic{table}}
\caption{Experimental conditions for all the FBG data obtained in the EMFC system.}
\vspace{+0.1cm}
\begin{tabular}{c|cccc} \hline\hline
~~Data~~ & ~~Shot~~ & ~~Sample position $z$~~ & ~~$B_{\rm max}$~~ & ~~Filter~~ \\ \hline
\#3 & I & 10~mm & 360~T & A \\
\#4 & II & 0~mm & 580~T & B \\
\#5 & I & 0~mm & 620~T & B \\
\#6 & II & 5~mm & 440~T & A \\ \hline\hline
\end{tabular}
\label{tab:S2}
\end{table}

\vspace{+0.8cm}

\newpage
\subsection*{\label{Sec8} \large Note 8. Analysis of the magnetostriction data obtained in the EMFC system}

We present the analysis method to subtract the extrinsic oscillation component from the experimental $\Delta L/L$ data.

\begin{figure}[h]
\centering
\includegraphics[width=0.9\linewidth]{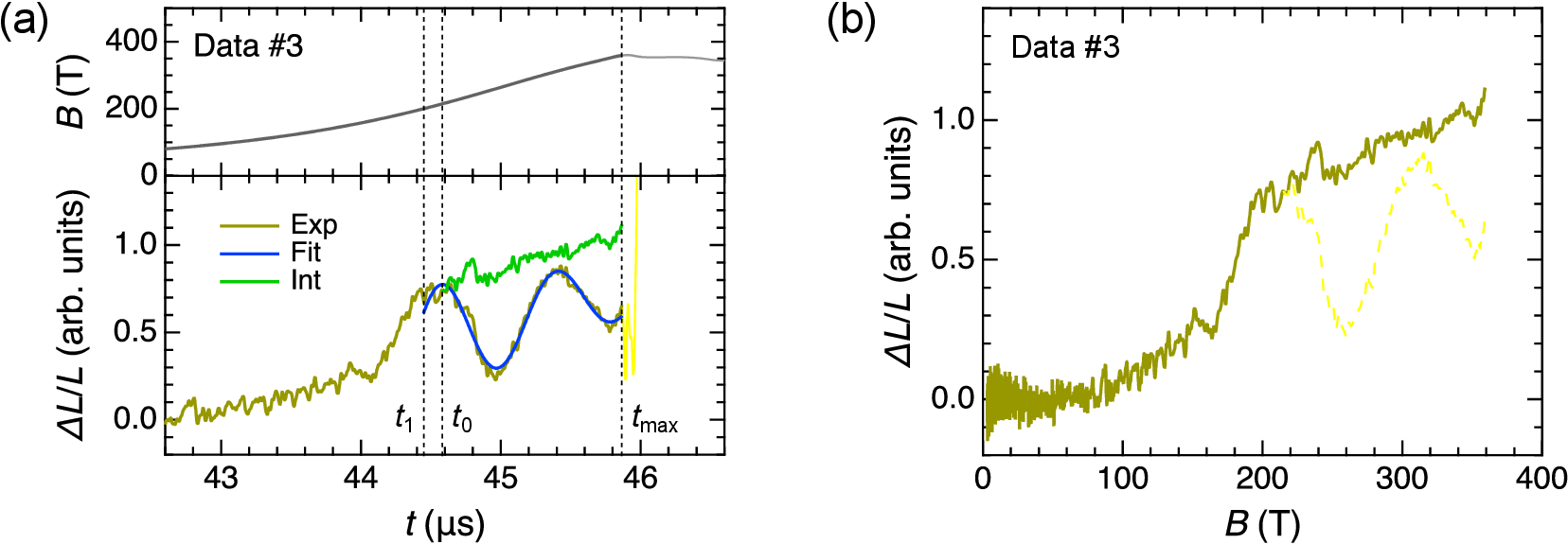}
\renewcommand{\thefigure}{S\arabic{figure}}
\caption{(a) Temporal evolutions of the magnetic field (upper panel) and the relative sample-length change $\Delta L/L$ (lower panel) obtained in the EMFC system for Data \#3. The blue line shows the fitting curve based on Eq.~(\ref{eq:S1}), and the green line shows the estimated intrinsic $\Delta L/L$ component. (b) $\Delta L/L$--$B$ curve before and after subtracting the extrinsic oscillation component for dashed and solid lines.}
\label{FigS10}
\end{figure}

We first analyze Data \#3 with $B_{\rm max} = 360$~T.
We fit the experimental $[\Delta L/L~(t)]_{\rm exp}$ curve using the following equation for the damped oscillation
\begin{equation}
\label{eq:S1}
[\Delta L/L~(t)]_{\rm fit} = \alpha + \beta (t - t_{0}) + A e^{-\gamma (t - t_{0})} \cos \omega (t - t_{0}),
\tag{S1}
\end{equation}
in a time range of $t_{1} \leq t \leq t_{\rm max}$, where $t_{1}$ and $t_{\rm max}$ are the time at which the magnetic field reaches 200~T and $B_{\rm max}$, respectively: $t_{1} = 44.448$~[$\mu$s], $t_{\rm max} = 45.868$~[$\mu$s].
The obtained fitting curve is shown by a blue line in Fig.~\ref{FigS10}(a).
The values of fitting parameters, $\alpha$, $\beta$, $t_{0}$, $A$, $\gamma$, and $\omega$, are summarized in Table~\ref{tab:S3}.
The agreement between $[\Delta L/L~(t)]_{\rm exp}$ and $[\Delta L/L~(t)]_{\rm fit}$ is good, indicating that the estimation of the oscillation component using the damped oscillation Eq.~(\ref{eq:S1}) is reasonable.
Accordingly, we can extract the intrinsic magnetostriction behavior $[\Delta L/L~(t)]_{\rm int}$ in $t_{0} \leq t \leq t_{\rm max}$ as shown by a green line in Fig.~\ref{FigS10}(a) using the following equation
\begin{equation}
\label{eq:S2}
[\Delta L/L~(t)]_{\rm int} = [\Delta L/L~(t)]_{\rm exp} - A [e^{-\gamma (t - t_{0})} \cos \omega (t - t_{0}) - 1].
\tag{S2}
\end{equation}
The resultant $\Delta L/L$--$B$ curve is shown in Fig.~\ref{FigS10}(b).

\begin{figure}[h]
\centering
\includegraphics[width=0.9\linewidth]{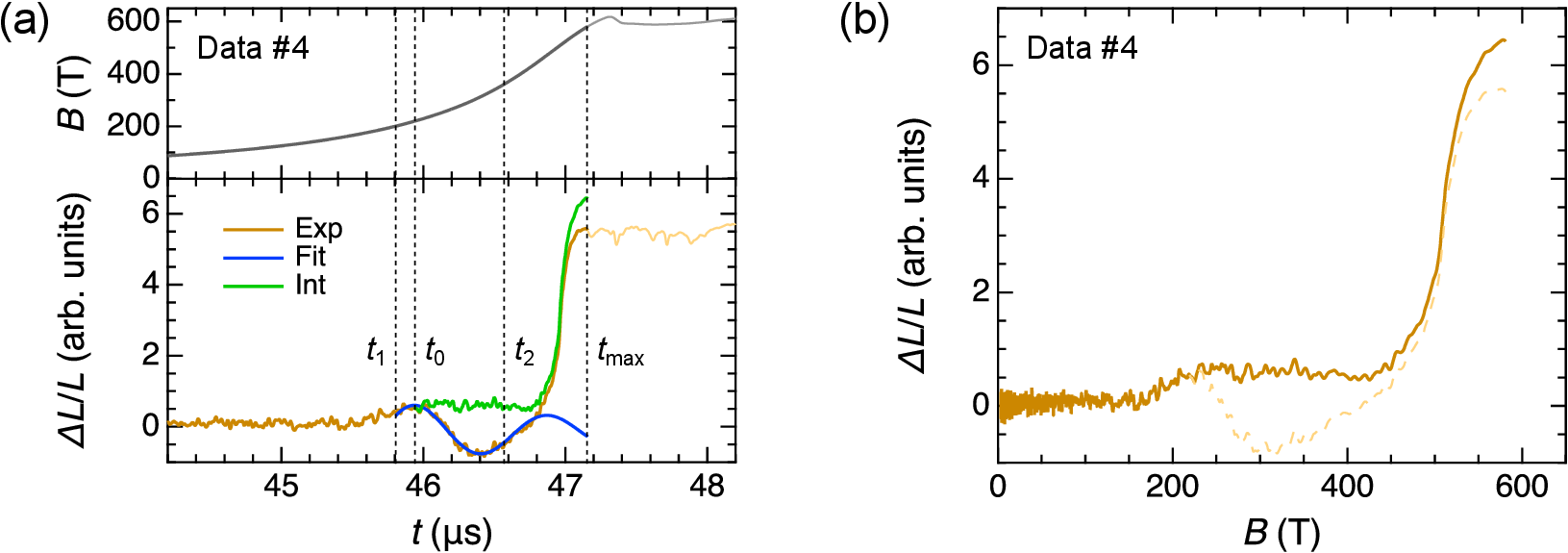}
\renewcommand{\thefigure}{S\arabic{figure}}
\caption{(a) Temporal evolutions of the magnetic field (upper panel) and the relative sample-length change $\Delta L/L$ (lower panel) obtained in the EMFC system for Data \#4 with $B_{\rm max} = 580$~T. The blue line shows the fitting curve based on Eq.~(\ref{eq:S1}), and the green line shows the estimated intrinsic $\Delta L/L$ component. (b) $\Delta L/L$--$B$ curve before and after subtracting the extrinsic oscillation component for dashed and solid lines.}
\label{FigS11}
\end{figure}

\begin{figure}[h]
\centering
\includegraphics[width=0.9\linewidth]{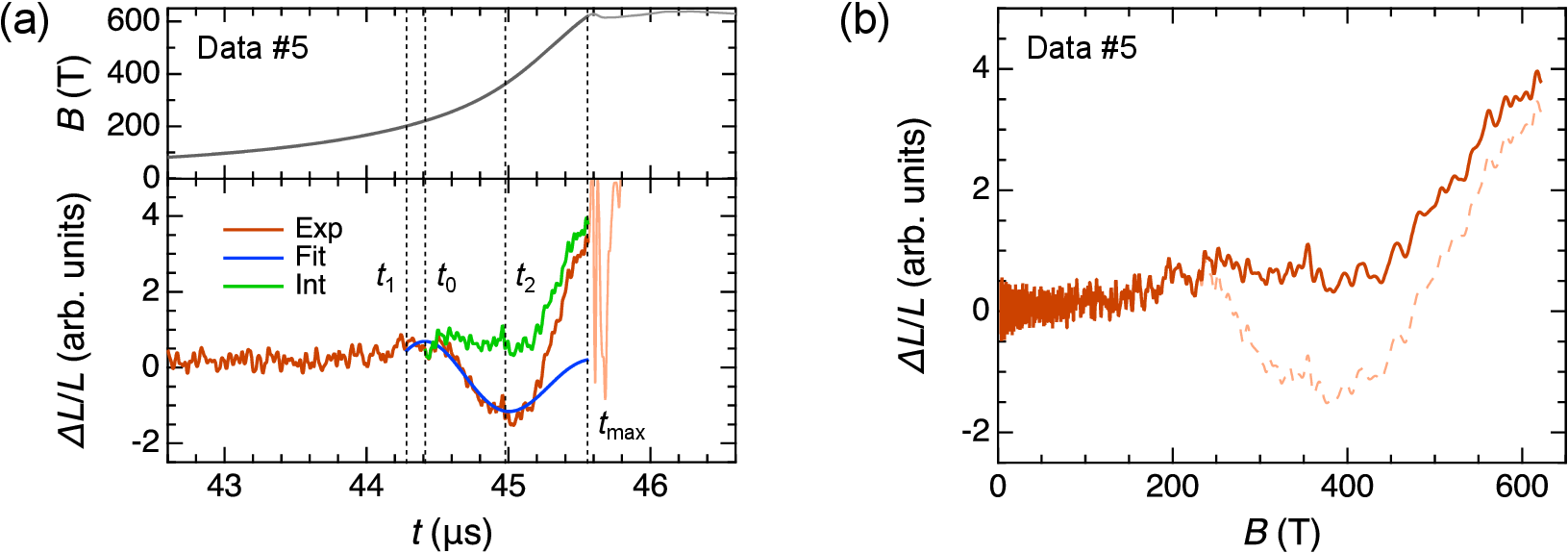}
\renewcommand{\thefigure}{S\arabic{figure}}
\caption{(a) Temporal evolutions of the magnetic field (upper panel) and the relative sample-length change $\Delta L/L$ (lower panel) obtained in the EMFC system for Data \#5 with $B_{\rm max} = 620$~T. The blue line shows the fitting curve based on Eq.~(\ref{eq:S1}), and the green line shows the estimated intrinsic $\Delta L/L$ component. (b) $\Delta L/L$--$B$ curve before and after subtracting the extrinsic oscillation component for dashed and solid lines.}
\label{FigS12}
\end{figure}

Based on the above analysis, we analyze Data \#4 and \#5 with $B_{\rm max} \sim 600$~T as shown below.
We adopt two $\Delta L/L$--$B$ curves shown in Figs.~\ref{FigS11}(b) and \ref{FigS12}(b) in the main text.
Note that we avoid detailed analysis for Data \#6 because we cannot obtain clear fitting results as well as it is hard to judge whether additional phase transition occurs in a high-field regime.

\vspace{+0.7cm}

\begin{table}[h]
\renewcommand{\arraystretch}{1.2}
\renewcommand{\thetable}{S\arabic{table}}
\caption{List of the fitting parameters in Eq.~(\ref{eq:S1}) adopted to the experimental $\Delta L/L$--$t$ curve in a time rage of $t_{1} \leq t \leq t_{\rm 2}$. Note that $t_{\rm 2} = t_{\rm max}$ for Data \#3 (see the text for details).}
\vspace{+0.1cm}
\begin{tabular}{c|cccccc} \hline\hline
~~Data~~ & ~~$\alpha$~~ & ~~$\beta$~[s$^{-1}$]~~ & ~~$t_{0}$~[$\mu$s]~~ & ~~$A$~~ & ~~$\gamma$~[s$^{-1}$]~~ & ~~$\omega$~[s$^{-1}$]~~\\ \hline
\#3 & ~~0.40~~ & ~~$2.19 \times 10^{5}$~~ & ~~44.584~~ & ~~0.311~~ & ~~$4.98 \times 10^{5}$~~ & ~~$7.64 \times 10^{6}$~~ \\
\#4 & ~~$-0.16$~~ & ~~0 (fix)~~ & ~~45.937~~ & ~~0.764~~ & ~~$4.98 \times 10^{5}$ (fix)~~ & ~~$6.73 \times 10^{6}$~~ \\
\#5 & ~~$-0.37$~~ & ~~0 (fix)~~ & ~~44.411~~ & ~~1.065~~ & ~~$4.98 \times 10^{5}$ (fix)~~ & ~~$5.30 \times 10^{6}$~~ \\ \hline\hline
\end{tabular}
\label{tab:S3}
\end{table}

\newpage
\hspace{-0.5cm}
[Data \#4]

Having confirmed that 1/2-magnetization plateau persists from 200~T to at least 360~T from Data \#3, we fit the experimental $[\Delta L/L~(t)]_{\rm exp}$ curve using Eq.~(\ref{eq:S1}) in a time range of $t_{1} \leq t \leq t_{\rm 2}$, where $t_{2}$ is the time at which the magnetic field reaches 360~T: $t_{1} = 45.804$~[$\mu$s], $t_{2} = 46.570$~[$\mu$s].
Here, we fix $\beta = 0$ because of the difficulty in its estimation and $\gamma = 4.98 \times 10^{5}$~[s$^{-1}$] following the value obtained for Data \#6.
The obtained fitting curve is shown by a blue line in Fig.~\ref{FigS11}(a), and the fitting parameters are summarized in Table~\ref{tab:S3}.
Finally, we extract the intrinsic magnetostriction behavior $[\Delta L/L~(t)]_{\rm int}$ in $t_{0} \leq t \leq t_{\rm max}$ as shown by a green line in Fig.~\ref{FigS11}(a) using Eq.~(\ref{eq:S2}).
The resultant $\Delta L/L$--$B$ curve is shown in Fig.~\ref{FigS11}(b).

\vspace{+0.5cm}

\vspace{+0.5cm}
\hspace{-0.5cm}
[Data \#5]

Similar to the analysis in Data \#4, we fit the experimental $[\Delta L/L~(t)]_{\rm exp}$ curve using Eq.~(\ref{eq:S1}) in a time range of $t_{1} \leq t \leq t_{\rm 2}$: $t_{1} = 44.276$~[$\mu$s], $t_{2} = 44.974$~[$\mu$s].
Here, we fix $\beta = 0$ because of the difficulty in its estimation and $\gamma = 4.98 \times 10^{5}$~[s$^{-1}$] following the value obtained for Data \#6.
The obtained fitting curve is shown by a blue line in Fig.~\ref{FigS12}(a), and the fitting parameters are summarized in Table~\ref{tab:S3}.
Finally, we extract the intrinsic magnetostriction behavior $[\Delta L/L~(t)]_{\rm int}$ in $t_{0} \leq t \leq t_{\rm max}$ as shown by a green line in Fig.~\ref{FigS12}(a) using Eq.~(\ref{eq:S2}).
The resultant $\Delta L/L$--$B$ curve is shown in Fig.~\ref{FigS12}(b).

\vspace{+0.5cm}

\newpage
\subsection*{\label{Sec9} \large Note 9. Site-phonon model on the breathing pyrochlore Heisenberg antiferromagnet}

In order to describe a microscopic magnetoelastic model, we start from the following spin Hamiltonian:
\begin{equation}
\label{H}
{\mathcal{H}}=\sum_{\langle i, j\rangle} J_{\rm ex} {\mathbf S}_{i} \cdot {\mathbf S}_{j}+\frac{c}{2}\sum_{i}{|{\mathbf u}_{i}|^{2}}-h\sum_{i}S_{i}^{z},
\tag{S3}
\end{equation}
where $\langle i, j\rangle$ runs over all the nearest-neighbor (NN) sites, $J_{\rm ex}~(>0)$ is the NN AFM exchange interaction, ${\mathbf S}_{i}$ is the classical spin at site $i$ normalized to $|{\mathbf S}_{i}|=1$, $c~(>0)$ is the elastic constant, ${\mathbf u}_{i}$ is the displacement at site $i$ from its original position ${\mathbf r}_{i}^{0}$, and $h$ is the strength of the external magnetic field applied along $z$ axis.
The exchange striction is introduced assuming $J_{\rm ex}$ linearly modulated by the bond-length change provided that $|{\mathbf u}_{i}|/|{\mathbf r}_{i}^{0}| \ll 1$:
\begin{equation}
\label{H_J}
J_{\rm ex} \equiv J_{\rm ex}(|{\mathbf r}_{ij}^{0}+{\mathbf u}_{i}-{\mathbf u}_{j}|) \approx J_{\rm ex}(|{\mathbf r}_{ij}^{0}|)+ \left .\frac{dJ_{\rm ex}}{dr}\right|_{r = |{\mathbf r}_{ij}^{0}|}{\mathbf e}_{ij} \cdot ({\mathbf u}_{i}-{\mathbf u}_{j}),
\tag{S4}
\end{equation}
where ${\mathbf r}_{ij}^{0} \equiv {\mathbf r}_{i}^{0}-{\mathbf r}_{j}^{0}$ and ${\mathbf e}_{ij} \equiv {\mathbf r}_{ij}^{0}/|{\mathbf r}_{ij}^{0}|$.
For the lattice degrees of freedom, we assume the so-called Einstein site phonons, where the displacements ${\mathbf u}_{i}$ are independent of each other \cite{2006_Ber, 2019_Aoy, 2021_Aoy}.
Substituting Eq.~(\ref{H_J}) to Eq.~(\ref{H}) and exactly integrating out the lattice degrees of freedom ${\mathbf u}_{i}$ using the standard Gaussian integration, we obtain
\begin{equation}
\label{H2}
{\mathcal{H}}=\sum_{\langle i, j\rangle} J_{\rm ex}(|{\mathbf r}_{ij}^{0}|) {\mathbf S}_{i} \cdot {\mathbf S}_{j}+\frac{c}{2}\sum_{i}{|{\mathbf u}_{i} - \bar{\mathbf u}_i|^{2}}-\frac{c}{2}\sum_{i}{|\bar{\mathbf u}_{i}|^{2}}-h\sum_{i}S_{i}^{z},
\vspace{-0.2cm}
\tag{S5}
\end{equation}
\begin{equation}
\label{H_u}
\bar{\mathbf u}_i = -\frac{1}{c} \sum_{j \in N(i)} \left .\frac{dJ_{\rm ex}}{dr}\right|_{r = |{\mathbf r}_{ij}^{0}|}{\mathbf e}_{ij} ({\mathbf S}_i \cdot {\mathbf S}_j).
\tag{S6}
\end{equation}
Obviously, the minimization condition is ${\mathbf u}_{i} = \bar{\mathbf u}_i$.

Here, we introduce two kinds of NN exchange interactions $J \equiv J_{\rm ex}(|{\mathbf r}_{ij}^{0}|_{A})$ and $J' \equiv J_{\rm ex}(|{\mathbf r}_{ij}^{0}|_{B})$ within tetrahedra {\it A} and {\it B}, respectively, in the breathing pyrochlore lattice [see Fig.~2(a) in the main text].
Then, the effective spin Hamiltonian can be expressed as 
\begin{equation}
\label{eq:H_eff}
{\mathcal{H}}_{\mathrm{eff}} = {\mathcal{H}}_{\mathrm{0}} + {\mathcal{H}}_{\mathrm{SLC}} - h\sum_{i}S^{i}
\tag{S7}
\end{equation}
with
\begin{equation}
\label{eq:H_0}
{\mathcal{H}_{0}}=J\sum_{\langle i, j\rangle_{A}}{\mathbf S}_{i} \cdot {\mathbf S}_{j}+J'\sum_{\langle i, j\rangle_{B}}{\mathbf S}_{i} \cdot {\mathbf S}_{j},
\vspace{-0.2cm}
\tag{S8}
\end{equation}
\begin{equation}
\label{eq:H_SLC_1}
{\mathcal{H}_{\mathrm{SLC}}}=-\frac{c}{2}\sum_{i}{|\bar{\mathbf u}_{i}|^{2}},
\vspace{-0.2cm}
\tag{S9}
\end{equation}
\begin{equation}
\label{H_u2}
\bar{\mathbf u}_i =  \left\{ \sqrt{\frac{Jb}{c}}\sum_{j\in N_{A}(i)} +  \sqrt{\frac{J'b'}{c}}\sum_{j\in N_{B}(i)} \right\} {\mathbf e}_{ij} ({\mathbf S}_i \cdot {\mathbf S}_j),
\tag{S10}
\end{equation}
where the dimensionless parameters $b$ ($b'$) represents the strength of the SLC between the NN sites within the tetrahedra {\it A} ({\it B}), which is defined by $b \equiv (1/cJ)[(dJ/dr)|_{r=|{\mathbf r}_{ij}^{0}|}]^{2}$ ($b' \equiv (1/cJ')[(dJ'/dr)|_{r=|{\mathbf r}_{ij}^{0}|}]^{2}$), and $N_{A}(i)$ ($N_{B}(i)$) denotes a set of the NN sites of site $i$ within tetrahedra {\it A} ({\it B}).
Expanding Eq.~(\ref{eq:H_SLC_1}), the SLC contribution can be eventually written as
\begin{equation}
\begin{split}
\label{eq:H_SLC_2}
{\mathcal{H}_{\mathrm{SLC}}}&=-Jb\sum_{\langle i, j\rangle_{A}}({\mathbf S}_{i} \cdot {\mathbf S}_{j})^2-J'b'\sum_{\langle i, j\rangle_{B}}({\mathbf S}_{i} \cdot {\mathbf S}_{j})^2\\
&\quad-\sum_{i} \left\{ \frac{Jb}{4}\sum_{j\neq k\in N_{A}(i)}+\frac{J'b'}{4}\sum_{j\neq k\in N_{B}(i)} \right\} ({\mathbf S}_{i} \cdot {\mathbf S}_{j})({\mathbf S}_{i}\cdot {\mathbf S}_{k})\\
&\quad-\sqrt{JJ'bb'}\sum_{i}\sum_{j\in N_{A}(i)}\sum_{k\in N_{B}(i)}{\mathbf e}_{ij} \cdot {\mathbf e}_{ik}({\mathbf S}_{i} \cdot {\mathbf S}_{j})({\mathbf S}_{i} \cdot {\mathbf S}_{k}),\\
\end{split}
\tag{S11}
\end{equation}
which is identical to Eq.~(3) in the main text.

\newpage
\subsection*{\label{Sec10} \large Note 10. System size dependence of the calculated results on Monte Carlo simulations}

\begin{figure}[h]
\centering
\includegraphics[width=0.9\linewidth]{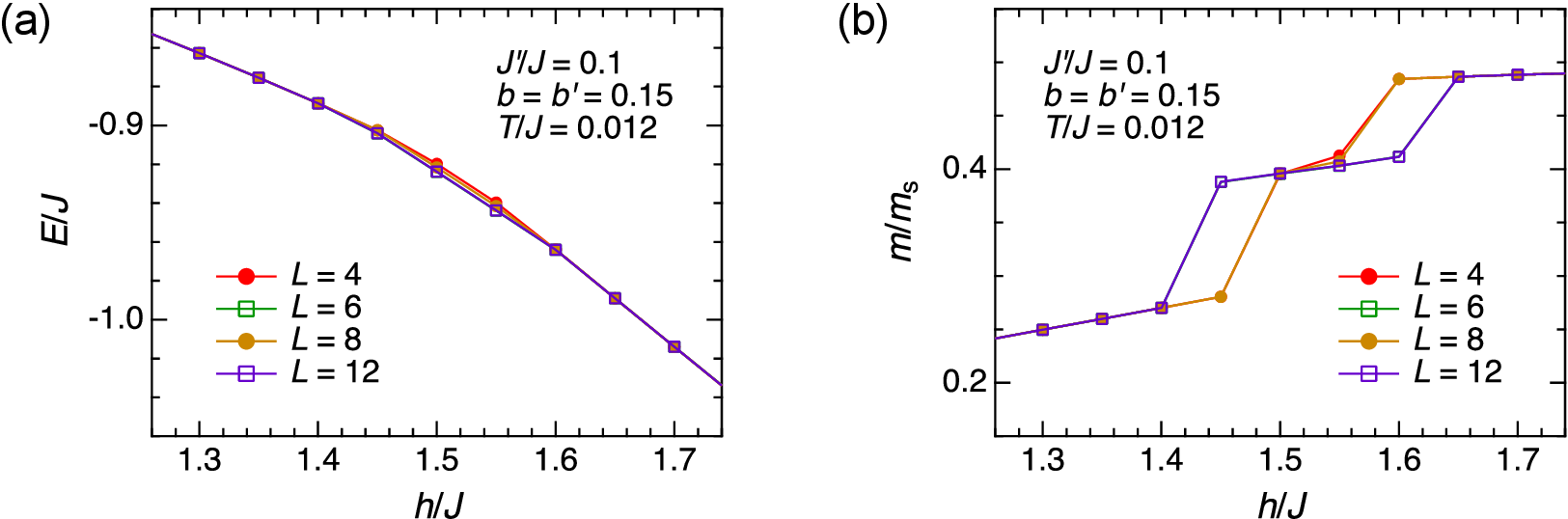}
\renewcommand{\thefigure}{S\arabic{figure}}
\caption{System size dependence of (a) the energy per site and (b) magnetization in the site-phonon model Eq.~(\ref{eq:H_eff}) for $J'/J = 0.1, b = b' = 0.015$, and $T/J = 0.012$. The data for $L = 6$ and $L = 12$ perfectly overlap with each other.}
\label{FigS13}
\end{figure}

In our Monte Carlo simulations on the site-phonon model Eq.~(\ref{eq:H_eff}) with $J'/J = 0.1$, we have checked the system-size $N = 16L^{3}$ dependence of the calculated results.
For $b = b' = 0.05$ and $b = b' = 0.10$, all the calculated results for $L = 4, 6, 8$ and 12 agree with each other.
For $b = b' = 0.15$, the energy per site in a field range of $1.4 < h/J < 1.6$ obtained for $L = 6$ and 12 is found to be smaller than that for $L = 4$ and 8 [Fig.~\ref{FigS13}(a)].
For $L = 6$ and 12, the same $6 \times 6 \times 6$ magnetic long-range order appears with a two-step metamagnetic transition [Fig.~\ref{FigS13}(b)].
For $L = 4$ and 8, on the other hand, we could not find any signatures of long-range orders with two-fold or four-fold periodicity.

\vspace{+0.8cm}

\newpage
\subsection*{\label{Sec11} \large Note 11. Magnetic-field width of the half-magnetization plateau}

A half-magnetization plateau commonly appears in Cr spinel oxides {\it A}Cr$_{2}$O$_{4}$ ({\it A} = Zn, Cd, and Hg) \cite{2011_Kim, 2013_Miy, 2011_Miy} as well as LiGaCr$_{4}$O$_{8}$.
Table~\ref{tab:S4} summarizes starting and ending fields of the half-magnetization plateau phase in these compounds.
Quantitative comparison of the values of $B_{\rm c3}/B_{\rm c2}$ between experiment and theory is useful for discussing the relative field width of the half-magnetization plateau.

\begin{table}[h]
\renewcommand{\arraystretch}{1.2}
\renewcommand{\thetable}{S\arabic{table}}
\caption{Starting and ending fields of the half-magnetization plateau, $B_{\rm c2}$ and $B_{\rm c3}$, respectively, in {\it A}Cr$_{2}$O$_{4}$ \cite{2011_Kim, 2013_Miy, 2011_Miy} and LiGaCr$_{4}$O$_{8}$. All these values were obtained at the lowest measured temperature.}
\vspace{+0.1cm}
\begin{tabular}{c|ccc} \hline\hline
~~~~ & ~~$B_{\rm c2}$~~ & ~~$B_{\rm c3}$~~ & ~~$B_{\rm c3}/B_{\rm c2}$~~ \\ \hline
~~ZnCr$_{2}$O$_{4}$~~ & ~~135~T~~ & ~~160~T~~ & ~~1.2~~ \\
~~CdCr$_{2}$O$_{4}$~~ & ~~28~T~~ & ~~58~T~~ & ~~2.1~~ \\
~~HgCr$_{2}$O$_{4}$~~ & ~~10~T~~ & ~~27~T~~ & ~~2.7~~ \\
~~LiGaCr$_{4}$O$_{8}$~~ & ~~$\sim$170~T~~ & ~~$\sim$420~T~~ & ~~2.5~~ \\ \hline\hline
\end{tabular}
\label{tab:S4}
\end{table}

\begin{figure}[h]
\centering
\includegraphics[width=0.9\linewidth]{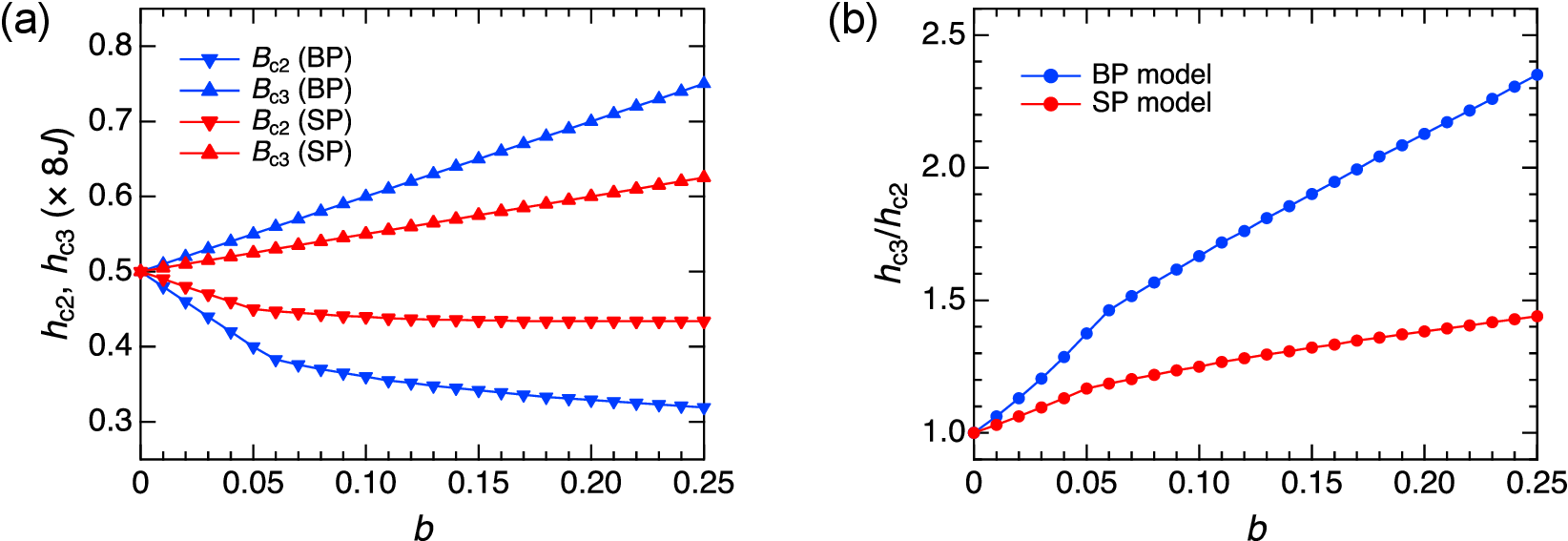}
\renewcommand{\thefigure}{S\arabic{figure}}
\caption{Spin-lattice-coupling parameter ($b = b'$) dependence of (a) starting and ending fields of the half-magnetization plateau, $h_{\rm c2}$ (downward triangles) and $h_{\rm c3}$ (upward triangles), respectively, and (b) the value of $h_{\rm c3}/h_{\rm c2}$ in the bond-phonon (BP, blue) and site-phonon (SP, red) models with $J'/J = 1$.}
\label{FigS14}
\end{figure}

\vspace{+0.5cm}
The ground-state phase diagrams of the bond-phonon and site-phonon models in a regular pyrochlore Heisenberg antiferromagnet ($J'/J = 1, b = b'$) were previously investigated in Ref.~\cite{2004_Pen} and Ref.~\cite{2021_Aoy}, respectively.
Note that, in the bond-phonon model, phonon-mediated spin interactions include only biquadratic terms identical to the first and second terms in Eq.~(\ref{eq:H_SLC_2}).
Figure~\ref{FigS14}(a) shows the spin-lattice-coupling parameter $b$ dependence of starting and ending fields of the half-magnetization $h_{\rm c2}$ and $h_{\rm c3}$, respectively, in the aforementioned two kinds of magnetoelastic models with $J'/J = 1$.
The increasing rate of $h_{\rm c3}$ with respect to $b$ in the site-phonon model is $dh_{\rm c3}/db = 0.5$, which is exactly half that in the bond-phonon model.
The decreasing rate of $h_{\rm c2}$ in the site-phonon model is also exactly half that in the bond-phonon model for $b < 0.05$ and becomes much smaller for $0.05 < b$ ($h_{\rm c2}$ is never less than $0.434 \times 8J$).
Consequently, the increase in $h_{\rm c3}/h_{\rm c2}$ with respect to $b$ is slower in the site-phonon model, as shown in Fig.~\ref{FigS14}(b).
As shown in Table~\ref{tab:S4}, $B_{\rm c3}/B_{\rm c2}$ amounts to more than 2 for CdCr$_{2}$O$_{4}$ and  HgCr$_{2}$O$_{4}$, which can be reproduced with the bond-phonon model, but not with the site-phonon model.
Even with the introduction of breathing anisotropy, i.e., $J'/J \neq 1$, the ground-state phase diagram of the bond-phonon model does not change from that for $J'/J = 1$ given that $b = b'$.
For the site-phonon model, $h_{\rm c3}/h_{\rm c2}$ becomes a bit larger than the case of $J'/J = 1$ but never exceeds 2 unlike the bond-phonon model: $h_{\rm c3}/h_{\rm c2} \approx 1.23$ for $b = b' = 0.05$, $h_{\rm c3}/h_{\rm c2} \approx 1.56$ for $b = b' = 0.10$, and $h_{\rm c3}/h_{\rm c2} \approx 1.63$ for $b = b' = 0.15$ (Figs.~5A--5C in the main text).
From the above, we conclude that the site-phonon model underestimates the field width of the half-magnetization plateau.

\vspace{+0.5cm}
